\documentclass[a4paper,11pt]{article}
\usepackage{draft}
\usepackage{cite}
\usepackage[T1]{fontenc}
\usepackage[utf8]{inputenc}
\usepackage{amsmath,amssymb,mathtools,bm}
\usepackage{booktabs,array,longtable}
\usepackage{enumitem}
\usepackage{orcidlink}
\usepackage{float}
\renewcommand{\afterTocRuleSpace}{\clearpage}

\definecolor{twilightlavender}{rgb}{0.54, 0.29, 0.42}
\definecolor{richmaroon}{rgb}{0.69, 0.19, 0.38}
\definecolor{forestgreen(web)}{rgb}{0.13, 0.55, 0.13}
\definecolor{lava}{rgb}{0.81, 0.06, 0.13}
\hypersetup{
	breaklinks,
	colorlinks,
	citecolor=forestgreen(web),
	filecolor=richmaroon,
	linkcolor=lava,
	urlcolor=twilightlavender
}
\usepackage{tikz}
\usetikzlibrary{arrows.meta,positioning,calc,fit,backgrounds}
\usepackage{tikz}
\usetikzlibrary{arrows.meta,calc}
\usepackage{tikz}
\usetikzlibrary{arrows.meta,positioning}
\usepackage{tikz}
\usetikzlibrary{
    arrows.meta,
    decorations.pathreplacing,
    calc,
    positioning
}
\usepackage[nameinlink,capitalise,noabbrev]{cleveref}
\usepackage{tikz}
\usetikzlibrary{arrows.meta,positioning,calc}
\crefformat{section}{\S#2#1#3} 
\crefformat{subsection}{\S#2#1#3}
\crefformat{subsubsection}{\S#2#1#3}

\numberwithin{equation}{section}
\allowdisplaybreaks

\newcommand{\AdS}{\mathrm{AdS}}
\newcommand{\Mink}{\mathbb{M}}
\newcommand{\cM}{\mathcal{M}}

\newcommand{\cL}{\mathcal{L}}
\newcommand{\cH}{\mathcal{H}}
\newcommand{\cS}{\mathcal{S}}

\newcommand{\dd}{\mathrm{d}}

\newcommand{\Lie}{\mathcal{L}}

\title{
{\sffamily\bfseries\fontsize{17}{21}\selectfont
Carroll-Weyl gauged null strings in AdS}
}

\affiliation[a]{Yau Mathematical Sciences Center (YMSC), Tsinghua University, Beijing 100084, China}
\affiliation[b]{Harish-Chandra Research Institute, A CI of Homi Bhabha National Institute,
Chhatnag Road, Jhunsi, Prayagraj (Allahabad), Uttar Pradesh 211019, India}

\author[a,\orcidlink{0000-0002-4535-3198}]{Sarthak Duary}\emailAdd{sarthakduary@tsinghua.edu.cn}
\author[b,\orcidlink{0009-0008-2963-2497}]{and Sourav Maji}\emailAdd{souravmaji@hri.res.in}

\abstract{We study the Carroll-Weyl gauged null string in the Poincar\'e patch of
$\mathrm{AdS}_D$ and its BRST quantization. In Poincar\'e coordinates
$X^A=(z,x^i)$, the radial conformal Killing vector $K=\partial_z$
becomes homothetic in a suitable conformally related frame. A
field-dependent rescaling of the homothetic Hamiltonian constraint
then leads to the radial-adapted basis
$\mathcal{M}=\frac{1}{2}(P_z^2+P_iP^i)$,
$\mathcal{L}=-P_AX'^A$, and
$\mathcal{S}=P_z$, with $\mathcal{M}$ invariant along the Carroll-Weyl
radial flow. Their modes form a $\mathrm{BMS}_3$ algebra
extended by an abelian weight-one current. We classify the central
extensions allowed by this residual algebra and find four independent
cohomology classes in the $LL$, $LM$, $LS$, and $SS$ channels. We then
construct the associated BRST complex, including the ghost pair for the
radial Carroll-Weyl constraint. In the canonical highest-weight
first-order representation, only the Virasoro-type central extension is
realized:
$(c_{LL},c_{LM},c_{LS},c_{SS})=(2D-54,0,0,0)$.
Cancellation of the corresponding nontrivial BRST cocycle therefore gives
$D=27$ in this representation. This differs structurally from the
minimal flat Carroll-Weyl theory, where the dilation constraint
$X\cdot P$ gives extra anomaly channels.}

\begin{document}
\maketitle

\section{Introduction}
\label{sec:introduction}

The tensionless limit of string theory provides a useful framework for studying the relation between worldsheet geometry, residual gauge symmetry, and the choice of quantum vacuum. A classical null string sweeps out a
degenerate worldsheet with a distinguished null direction and an intrinsic
Carrollian geometry.  The constructions of Schild and of
Isberg-Lindstr\"om-Sundborg-Theodoridis (ILST) provide the standard
Lagrangian and Hamiltonian descriptions of such strings
\cite{Schild1977,IsbergLindstromSundborg1992,IsbergEtAl1994}.  After an
appropriate gauge fixing, the two conventional null-string constraints
generate the two-dimensional conformal Carroll algebra, isomorphic to
$\mathrm{BMS}_3$, which may also be obtained from a suitable contraction of
the tensile-string constraint algebra \cite{BagchiWorldsheet2015}.  At the
quantum level, however, the classical theory admits inequivalent vacuum
prescriptions \cite{BagchiThreeVacua2020,BagchiEtAlReview}.  This
sensitivity to quantization is also important in the relation between null
and ambitwistor strings \cite{MasonSkinner2014,CasaliTourkine2016}, and in
recent developments concerning open null strings and scattering in the
flipped vacuum \cite{BagchiEtAlOpenNull,BagchiEtAlScattering}.

A separate question concerns whether the conventional two-constraint
description captures the full local gauge redundancy of the null-string
worldsheet.  Sheikh-Jabbari and Yavartanoo argued that the degenerate
worldsheet geometry possesses an additional local Carroll-Weyl symmetry
which should itself be gauged
\cite{SheikhJabbariYavartanooOverlooked,SheikhJabbariYavartanooI,SheikhJabbariYavartanooClassical}.
The corresponding Hamiltonian formulation contains a third first-class
constraint in addition to the two ILST constraints
\cite{SheikhJabbariYavartanooII}, while the geometric distinction between
the Carroll-Weyl gauged and ILST descriptions has been further clarified
in ref.~\cite{SheikhJabbariYavartanooClassical}. Carroll-Weyl transformations have also recently appeared in flat holography, where they arise as asymptotic symmetries associated with the Carrollian geometry of null infinity and exhibit a nontrivial anomaly structure \cite{Delfante:2026ads}.

In the minimal Cartesian realization in flat target space, the three
constraints take the form
\begin{equation}
 C_1=P^2,
 \qquad
 C_2=P\cdot X',
 \qquad
 C_3=X\cdot P .
 \label{eq:flatconstraints}
\end{equation}
The third constraint generates local target-space dilations and identifies
configurations along the corresponding gauge orbit.  On a regular
constraint stratum this reduces the number of local configuration degrees
of freedom from $(D-2)$ to $(D-3)$.  A complementary geometric interpretation
was given by Lindstr\"om and Sundborg, where fixing the local scale through
an embedding-space condition reduces the Carroll-Weyl system to an ILST
theory on a lower-dimensional (A)dS target
\cite{LindstromSundborg}.  We shall therefore distinguish between the
unreduced Carroll-Weyl theory, with its three first-class constraints,
and the reduced ILST description obtained after eliminating the additional
gauge direction.

The curved-background extension of this construction was developed in
ref.~\cite{NullHolography}.  The Carroll-Weyl flow is then associated with
a distinguished target-space vector field.  For the minimal construction
this vector is homothetic, while a more general formulation can be based
on a conformal Killing vector and related locally to the homothetic
description by a change of conformal frame.  On a regular non-null patch,
eliminating the Carroll-Weyl gauge direction gives a null string theory on
the corresponding local space of gauge orbits.

The Poincar\'e patch of $\AdS_D$ provides a particularly simple realization
of this construction.  The radial vector field $ K=\partial_z$ is a conformal Killing vector of the physical AdS metric.  After passing to
a suitable conformally related metric, the same flow becomes a normalized
homothety and can be gauged in the minimal Carroll-Weyl form.  The radial
coordinate then parametrizes the local gauge orbit, while the
$(D-1)$ coordinates $x^i$ remain invariant.  Classically, eliminating this
radial gauge direction gives an ILST theory on $\Mink^{D-1}$ \cite{NullHolography}.

The radial-adapted constraint basis makes the structure of this quotient
transparent already in the unreduced Carroll-Weyl theory. In the
homothetic frame, the natural Hamiltonian constraint
$\cM_g=\tfrac12 g^{AB}P_AP_B$ contains a $z$-dependent conformal factor.
On a regular patch, multiplication by the nonvanishing factor
$F=L^2e^{2z}$ gives
\begin{equation}
 \cM_{\rm rad}=F\cM_g
 =\frac12(P_z^2+P_iP^i),
\end{equation}
which is invariant along the radial gauge flow and defines the same local
classical constraint system. We use this radial-adapted representative in our analysis.

The quantum problem, however, cannot be inferred from the classical
reduction alone.  The complete gauge complex, including the additional
ghost sector, the residual constraint algebra, its possible central
extensions, and the choice of quantum representation must all be taken
into account.  This issue is already visible in the flat Cartesian
Carroll-Weyl theory.  The additional Faddeev-Popov sector was derived in
ref.~\cite{DuaryMajiPI}, and the corresponding highest-weight BRST
calculation gives
\begin{equation}
 (c_{LL},c_{LS},c_{SS})_{\rm flat}
 =
 \bigl(2D-54,\,6-D,\,4-D\bigr)
 \label{eq:flatvector}
\end{equation}
in the dilation realization \cite{DuaryMajiBRST}.  The three independent
conditions would require $D=27$, $D=6$, and $D=4$, respectively, and hence
cannot be satisfied simultaneously.  Light-cone and induced-vacuum
quantizations probe different realizations of the null-string quantum
theory \cite{RasulianEtAlLightcone,ChenHuAnomalies}.  The anomaly structure
of the AdS radial system must therefore be determined independently rather
than inherited from the flat Cartesian theory.

The other crucial ingredient is the form of the third constraint. For a
Carroll-Weyl flow generated by a target-space vector field
$K=K^A(X)\partial_A$, the corresponding canonical generator is
$K^AP_A$. In the AdS Poincar\'e construction this gives
\begin{equation}
 \cS=P_z.
 \label{eq:Spz}
\end{equation}
Together with the radial-adapted Hamiltonian representative introduced
above, this gives
\begin{equation}
 [S_m,M_n]=0.
\end{equation}
The natural homothetic constraint instead satisfies
$\{\cS,\cM_g\}=2\cM_g$.
Thus the vanishing $SM$ bracket refers specifically to the radial-adapted
constraint system used here.

The resulting centerless residual algebra consists of the usual
$\mathrm{BMS}_3$ sector generated by $(L_m,M_m)$ together with an abelian
weight-one current $S_m$ commuting with the supertranslation sector.
Closely related one-current extensions of $\mathrm{BMS}_3$ have appeared
in refs.~\cite{Hao:2021BMS,Yu:2022BMSIsing,Liu:2026LoopBMSKM}; in the
present construction the additional current has a direct geometric origin
as the constraint associated with the gauged AdS radial flow.

This distinction also changes the possible quantum anomaly structure.
It is useful here to separate two questions.  The first is purely
algebraic: which central extensions are permitted by the residual Lie
algebra?  The second is representational: which of these allowed classes
are actually generated by a particular normal-ordered matter-ghost
system?  An explicit Chevalley-Eilenberg analysis of the AdS radial
algebra gives
\begin{equation}
 H^2(\mathfrak a_{\rm AdS};\mathbb C)
 \simeq
 \mathbb C^4,
\end{equation}
with independent representatives in the $LL$, $LM$, $LS$, and $SS$
channels.  There are no independent $MM$ or $MS$ cocycles.  The existence
of the $LM$ class is particularly noteworthy: because $[S_m,M_n]=0$, the
usual mixed central extension of the $\mathrm{BMS}_3$ sector is compatible
with the full radial algebra, unlike in the flat dilation realization.

We then quantize this three-constraint system in a specified
highest-weight first-order representation.  The matter currents are
\begin{equation}
 T^X=-:P_A\partial X^A:,
 \qquad
 M^X=\frac12:P_AP^A:,
 \qquad
 S^X=P_z,
\end{equation}
and each first-class constraint is accompanied by its corresponding
fermionic ghost--antighost pair.  The structure constants of the radial
algebra determine the semidirect ghost currents and the cubic terms in the
BRST charge.  In particular, the vanishing of $[S_m,M_n]$ removes the
ghost interaction that is present in the flat dilation complex.

\paragraph{Main results.}
For the radial Carroll-Weyl flow in the AdS Poincar\'e patch, we obtain
the three first-class constraints
\begin{equation}
 \cM=\frac12(P_z^2+P_iP^i),
 \qquad
 \cL=-P_AX'^A,
 \qquad
 \cS=P_z .
\end{equation}
Their modes form a $\mathrm{BMS}_3$ algebra extended by an abelian
weight-one current $S_m$, with $[S_m,M_n]=0$.  An explicit
Chevalley-Eilenberg analysis shows that this algebra admits four
independent central classes in the $LL$, $LM$, $LS$, and $SS$ channels.

Quantizing the system in the canonical highest-weight first-order
representation, including the ghost pair associated with $\cS$, we find
\begin{equation}
 (c_{LL},c_{LM},c_{LS},c_{SS})_{\rm tot}
 =
 (2D-54,0,0,0).
 \label{eq:intro-total-anomaly}
\end{equation}
Thus, although four central extensions are allowed algebraically, only the
Virasoro-type cocycle is realized in this representation.  Cancellation of
the corresponding BRST obstruction gives $D=27$.

The difference from the flat Carroll-Weyl no-go result reflects both the
radial generator $\cS=P_z$ and the radial-adapted Hamiltonian
representative used here. Together they give $[S_m,M_n]=0$, changing the
residual algebra and the corresponding ghost couplings and eliminating
the $LS$ and $SS$ anomaly channels in the canonical AdS realization. The relation to the $(D-1)$-dimensional reduced ILST theory is developed
in detail in the companion paper~\cite{Duary:2026izp}.
There, starting from the radial-adapted formulation used here, a further
quotient-adapted BFV basis separates the transverse ILST complex from a
contractible radial quartet. An explicit contracting homotopy on the
local jet algebra shows that the radial sector is BRST trivial, while the
same quartet carries zero net Virasoro central charge and drops out of the
regulated canonical functional measure. At the anomaly-free point this
gives the dimensional descent $D=27\rightarrow d=26$. The radial-adapted
basis and the quotient-adapted BFV basis are therefore successive,
distinct steps. In the present work we focus on the unreduced AdS
Carroll-Weyl constraint system, its residual algebra, and its BRST
quantization; extending these results to an independently quantized
unrescaled homothetic constraint requires the corresponding quantum BFV
equivalence.
\paragraph{Organization of the paper.}
The rest of the paper is organized as follows.
\cref{sec:curved_backgrounds} reviews the Carroll-Weyl construction
on curved backgrounds. \cref{sec:ads-poincare} specializes it
to the AdS Poincar\'e patch and derives the radial constraint algebra and
its central extensions. \cref{sec:brst-anomaly} constructs the
BRST complex, determines the quantum anomaly coefficients, and compares
the result with the flat Carroll-Weyl theory. \cref{app:Killing-homothety-CKV} reviews the
relevant target-space geometry. \cref{app:radial-constraint-checks} verifies the radial constraint
canonically. \cref{app:ads-residual-algebra} derives the residual mode algebra. \cref{app:adscohomology}
contains the Chevalley-Eilenberg analysis of its central extensions, and \cref{app:ope-details} gives the OPE and BRST-anomaly details.

\section{Carroll-Weyl null strings on curved backgrounds}
\label{sec:curved_backgrounds}

The Carroll-Weyl formulation of null strings is most transparent in flat
target space, where the local Carroll-Weyl transformation acts as a rescaling
of the embedding coordinates. Extending this construction to a curved target
space requires some care. In contrast with the ordinary Weyl symmetry of the
tensile Polyakov string, the Carroll-Weyl transformation acts nontrivially on
the target-space coordinates themselves. Consequently, simply replacing the
Minkowski metric by a generic curved metric does not in general preserve the
Carroll-Weyl gauge symmetry \cite{NullHolography}.

The appropriate curved-space formulation therefore depends on the existence
of a distinguished target-space vector field generating the Carroll-Weyl
flow. We first describe the minimal construction, for which this vector field
is required to be homothetic, and then discuss a broader class of backgrounds
admitting a conformal Killing vector. For completeness, we collect in~\cref{app:Killing-homothety-CKV} the geometric definitions and basic properties of Killing, homothetic and conformal Killing vector fields, together with the examples most relevant for the present construction.

\subsection{Minimal null strings on non-flat backgrounds}
\label{subsec:minimal_curved}

Let $(\mathcal{M},g)$ be a $D$-dimensional target manifold equipped with a
metric $g_{\mu\nu}$. The minimal Carroll-Weyl construction assumes that, at
least locally, the target space admits a homothetic vector field $\rho^\mu$
satisfying $  \nabla_\mu \rho_\nu+\nabla_\nu \rho_\mu
    =
    2g_{\mu\nu}$.
The normalization fixes the parametrization
along the homothetic flow.

The Carroll-Weyl covariant derivative of the embedding coordinates is then
defined by
\begin{equation}
    D_aX^\mu
    =
    \partial_aX^\mu+W_a\rho^\mu(X),
    \label{eq:CW_covariant_derivative_curved}
\end{equation}
where $W_a$ is the Carroll-Weyl gauge connection. The curved-space minimal
null-string action is
\begin{equation}
    S_{\rm MNS}[g;V,W]
    =
    \frac{\kappa}{2}
    \int_{\Sigma} d^2\sigma\,
    V^aV^b
    g_{\mu\nu}(X)
    D_aX^\mu D_bX^\nu .
    \label{eq:MNS_curved_action}
\end{equation}
Here $V^a$ is the worldsheet vector density characterizing the degenerate
null worldsheet geometry.

The local Carroll-Weyl transformations take the form
\begin{align}
    \delta_\chi X^\mu
    &=
    \chi\,\rho^\mu(X),
    \label{eq:CW_transform_X_curved}
    \\
    \delta_\chi W_a
    &=
    -\partial_a\chi,
    \label{eq:CW_transform_W_curved}
    \\
    \delta_\chi V^a
    &=
    -\chi V^a.
    \label{eq:CW_transform_V_curved}
\end{align}
Thus the transformation of the embedding coordinates is generated by the
homothetic flow of the target space. This is the curved-space generalization
of the flat-space transformation
\begin{equation}
    \delta_\chi X^\mu=\chi X^\mu.
\end{equation}
Indeed, in flat Minkowski space we take $  \rho^\mu=X^\mu$, for which the Euler vector $ \rho=X^\mu\partial_\mu$ is homothetic.

Under \eqref{eq:CW_transform_X_curved}--\eqref{eq:CW_transform_W_curved},
the covariant derivative transforms as
\begin{equation}
    \delta_\chi(D_aX^\mu)
    =
    \chi\,(\partial_\nu\rho^\mu)D_aX^\nu.
    \label{eq:CW_covariant_derivative_transformation}
\end{equation}
Together with
\begin{equation}
    \mathcal{L}_{\rho}g_{\mu\nu}=2g_{\mu\nu},
\end{equation}
and the transformation of $V^a$, this guarantees the invariance of
\eqref{eq:MNS_curved_action} under the local Carroll-Weyl symmetry.

This construction makes clear why a generic curved metric does not
automatically provide an admissible background for the minimal
Carroll-Weyl null string. Since the gauge transformation moves the embedding
coordinates along a target-space vector field, the target metric must
transform in a controlled way along this flow. In the minimal construction,
this condition is precisely the existence of a homothetic vector field.

It is useful to introduce coordinates adapted to the homothetic flow. On a
patch where $\rho^\mu$ is nonvanishing, we choose coordinates $ X^\mu=(s,Y^i)$,
such that $ \rho^\mu\partial_\mu=\partial_s$.
On a patch where the homothetic vector is non-null, we can further choose the normalization $   \rho^2=\pm e^{2s}$. The homothety condition then implies that the metric can be written as
\begin{equation}
    g_{\mu\nu}(s,Y)
    =
    e^{2s}h_{\mu\nu}(Y),
    \qquad
    \partial_s h_{\mu\nu}=0.
    \label{eq:homothetic_metric_decomposition}
\end{equation}

In the adapted coordinates, the Carroll-Weyl transformations become
\begin{equation}
    \delta_\chi s=\chi,
    \qquad
    \delta_\chi Y^i=0.
    \label{eq:CW_adapted_coordinates}
\end{equation}
It is therefore natural to introduce the Carroll-Weyl invariant combinations $ U^a=e^sV^a,$
and  $B_a=\partial_as+W_a$.
Indeed,
\begin{equation}
    \delta_\chi U^a=0,
    \qquad
    \delta_\chi B_a=0.
\end{equation}
The coordinate $s$ therefore parametrizes the Carroll-Weyl gauge orbit,
while the coordinates $Y^i$ are invariant under the local symmetry.

The same orbit-adapted viewpoint extends to the Hamiltonian constraints.
For $\cS=P_s$ and the metric
\eqref{eq:homothetic_metric_decomposition},
\[
 \cM_g=\frac12e^{-2s}h^{\mu\nu}(Y)P_\mu P_\nu,
 \qquad
 \cM_0=e^{2s}\cM_g
 =\frac12h^{\mu\nu}(Y)P_\mu P_\nu,
 \qquad
 \{P_s,\cM_0\}=0.
\]
Since $e^{2s}$ is nonvanishing on the patch, $\cM_g$ and $\cM_0$
define the same local classical constraint system, while $\cM_0$ is
invariant along the homothetic gauge flow. This general construction
motivates the radial-adapted Hamiltonian representative introduced below
for the AdS Poincar\'e patch; it does not remove the need to distinguish
between non-null and null homothetic directions.

An important distinction concerns the causal character of the homothetic
vector. Its norm $ \rho^2=g_{\mu\nu}\rho^\mu\rho^\nu$
need not have the same sign throughout the full target space. Even if a
homothety is defined globally, it can be spacelike in one region, timelike in
another, and null on a separating locus. The construction should therefore
in general be understood patchwise, with each patch chosen such that the
causal character of $\rho^\mu$ does not change.

The cases $\rho^2\neq0$ and $\rho^2=0$ are structurally different. To see
this, define
\begin{equation}
    v^\mu
    :=
    V^a\partial_aX^\mu,
    \qquad
    w:=V^aW_a.
\end{equation}
Then the action contains the combination $ g_{\mu\nu}v^\mu v^\nu
    +
    2w\,\rho_\mu v^\mu
    +
    w^2\rho^2.$
For $\rho^2\neq0$, the dependence on $w$ is quadratic, and its equation of
motion determines
\begin{equation}
    w
    =
    -\frac{\rho_\mu v^\mu}{\rho^2}.
    \label{eq:w_nonnull}
\end{equation}
For $\rho^2=0$, however, the quadratic term disappears. The gauge field then
acts as a Lagrange multiplier whose equation of motion imposes
\begin{equation}
    \rho_\mu V^a\partial_aX^\mu=0.
    \label{eq:null_homothety_constraint}
\end{equation}
Thus the null homothetic case cannot be obtained simply by taking the
$\rho^2\rightarrow0$ limit of the non-null construction and must be treated
separately.

The requirement of a genuine homothety is rather restrictive. In particular,
many backgrounds of physical interest do not admit homothetic vector fields
in their physical metric. This motivates a generalization of the above
construction to backgrounds admitting conformal Killing vectors.

\subsection{Nearest-to-minimal null-string backgrounds}
\label{subsec:nearest_minimal}

Let $(\mathcal{M},G)$ be a $D$-dimensional target space admitting a conformal
Killing vector $K^\mu$. The conformal Killing equation is
\begin{equation}
    \mathcal{L}_K G_{\mu\nu}
    =
    2\omega_K(X)G_{\mu\nu},
    \label{eq:CKV_equation}
\end{equation}
where
\begin{equation}
    \omega_K(X)
    =
    \frac{1}{D}\nabla_\mu K^\mu.
    \label{eq:omega_K}
\end{equation}
The special cases $\omega_K=0$ and $\omega_K=1$
correspond respectively to an ordinary Killing vector and a normalized homothetic
vector. A generic conformal Killing vector has a nonconstant
$\omega_K(X)$.

The flow generated by $K^\mu$ can be gauged by defining
\begin{equation}
    D_aX^\mu
    =
    \partial_aX^\mu+W_aK^\mu(X).
    \label{eq:NMNS_covariant_derivative}
\end{equation}
The corresponding action is
\begin{equation}
    S_{\rm NMNS}[G;E,W;K]
    =
    \frac{\kappa}{2}
    \int_{\Sigma}d^2\sigma\,
    E^aE^b
    G_{\mu\nu}(X)
    D_aX^\mu D_bX^\nu.
    \label{eq:NMNS_action}
\end{equation}
The local gauge transformations are
\begin{align}
    \delta_\chi X^\mu
    &=
    \chi K^\mu(X),
    \label{eq:NMNS_transform_X}
    \\
    \delta_\chi W_a
    &=
    -\partial_a\chi,
    \label{eq:NMNS_transform_W}
    \\
    \delta_\chi E^a
    &=
    -\chi\omega_K(X)E^a.
    \label{eq:NMNS_transform_E}
\end{align}
Using \eqref{eq:CKV_equation}, one verifies that
\eqref{eq:NMNS_action} is invariant under these transformations.

The difference from the minimal theory is important. For a homothety,
$\omega_K=1$, the transformation of the worldsheet vector density is
\begin{equation}
    \delta_\chi E^a=-\chi E^a,
\end{equation}
which depends only on the worldsheet gauge parameter. For a generic conformal
Killing vector, however, $\omega_K$ is a function of the target-space
coordinates, so that the transformation of $E^a$ explicitly depends on the
embedding fields $X^\mu$. The Carroll-Weyl transformation of the worldsheet
geometric variable therefore becomes target-space dependent.

For this reason, the construction based on a generic conformal Killing vector
is referred to as a nearest-to-minimal null-string theory rather than a
strictly minimal one.

The connection between the two constructions can nevertheless be made
explicit locally. Consider the conformally related target-space metric
\begin{equation}
    g_{\mu\nu}
    =
    e^{-2\Omega(X)}G_{\mu\nu}.
    \label{eq:conformal_metric_redefinition}
\end{equation}
Its Lie derivative along $K$ is
\begin{equation}
    \mathcal{L}_K g_{\mu\nu}
    =
    2\left(
        \omega_K
        -
        K^\rho\partial_\rho\Omega
    \right)
    g_{\mu\nu}.
    \label{eq:Lie_conformal_metric}
\end{equation}
If $\Omega(X)$ is chosen to satisfy
\begin{equation}
    K^\rho\partial_\rho\Omega
    =
    \omega_K-1,
    \label{eq:Omega_condition}
\end{equation}
then
\begin{equation}
    \mathcal{L}_K g_{\mu\nu}
    =
    2g_{\mu\nu}.
    \label{eq:K_becomes_homothety}
\end{equation}
Thus $K^\mu$ becomes a homothetic vector field of the conformally related
metric $g_{\mu\nu}$.

On a patch where $K^\mu$ is nonvanishing, the first-order equation
\eqref{eq:Omega_condition} can be integrated locally along the integral curves
of $K^\mu$. The existence of a globally smooth solution is not guaranteed,
and the construction is therefore naturally patchwise.

At the same time, define the rescaled worldsheet vector density
\begin{equation}
    V_K^a
    =
    e^{\Omega(X)}E^a.
    \label{eq:VK_definition}
\end{equation}
Using \eqref{eq:NMNS_transform_E} and
\eqref{eq:Omega_condition}, one obtains
\begin{equation}
    \delta_\chi V_K^a
    =
    -\chi V_K^a.
    \label{eq:VK_transformation}
\end{equation}
Moreover,
\begin{equation}
    E^aE^bG_{\mu\nu}
    =
    V_K^aV_K^b g_{\mu\nu}.
    \label{eq:metric_vector_relation}
\end{equation}
Consequently, the nearest-to-minimal action can be rewritten as
\begin{equation}
    S_{\rm NMNS}[G;E,W;K]
    =
    S_{\rm MNS}[g;V_K,W],
    \label{eq:NMNS_MNS_equivalence}
\end{equation}
where $K^\mu$ is now a homothetic vector of the metric $g_{\mu\nu}$.

Thus, locally, a Carroll-Weyl theory associated with a generic conformal
Killing vector can be mapped to the minimal construction in an appropriate
conformal frame. This considerably enlarges the class of backgrounds on
which the Carroll-Weyl null-string theory can be formulated.

There is, however, an additional choice involved. A given target space may
admit several inequivalent conformal Killing vectors. Different choices of
$K^\mu$ define different Carroll-Weyl gauge flows and need not lead to
equivalent realizations of the theory. The specification of a
nearest-to-minimal null-string background therefore includes not only the
target-space geometry but also the particular conformal Killing vector that
is being gauged.

The causal character of $K^\mu$ is likewise important. A given conformal
Killing vector can be spacelike, timelike or null in different regions of the
target manifold, and its norm can change sign across the spacetime. The
corresponding Carroll-Weyl construction should therefore be formulated on
patches in which $K^\mu$ remains nonvanishing and has a definite causal
character.

This framework is particularly well suited to the AdS Poincar\'e patch
considered below. Although the physical AdS metric does not admit a
homothetic vector of the type required by the minimal construction,
suitable conformal Killing vectors can be used to define a
nearest-to-minimal Carroll-Weyl null-string theory. In an appropriate
conformal frame, the selected conformal Killing direction becomes
homothetic, providing the starting point for the AdS construction of the
next section.

\section{Carroll-Weyl null strings in the AdS Poincar\'e patch}
\label{sec:ads-poincare}

We now specialize the general curved-background construction of
\cref{subsec:nearest_minimal} to $D$-dimensional AdS space.
Since the Carroll-Weyl gauging is organized around a distinguished
target-space flow, the Poincar\'e patch provides a particularly convenient
setting.  We begin by recalling the relevant geometry of this patch, then
identify the radial conformal Killing vector and the associated homothetic
frame.  This will provide the geometric starting point for constructing the
Carroll-Weyl action, deriving its phase-space constraints, and determining
the resulting residual symmetry algebra and its possible central extensions.

\subsection{Geometry of the AdS Poincar\'e patch}
\label{subsec:ads-preliminaries}

Before turning to the Carroll-Weyl construction, we briefly recall the
geometric relation between global AdS space and its Poincar\'e
patch.  This will also clarify the distinguished role played by the radial
coordinate $z$ in the analysis below.

Anti-de Sitter space $\AdS_D$ is realized as the hyperboloid
\begin{equation}
 -(X^{-1})^2-(X^0)^2
 +\sum_{A=1}^{D-1}(X^A)^2
 =
 -L^2
 \label{eq:ads-embedding}
\end{equation}
embedded in $\mathbb{R}^{2,D-1}$ with metric
\begin{equation}
 \dd s_{\rm emb}^2
 =
 -(\dd X^{-1})^2-(\dd X^0)^2
 +\sum_{A=1}^{D-1}(\dd X^A)^2 ,
\end{equation}
where $L$ is the AdS curvature radius.

A convenient global parametrization is
\begin{align}
 X^{-1}
 &=
 L\cosh\rho\,\cos\tau,
 &
 X^0
 &=
 L\cosh\rho\,\sin\tau,
 \nonumber\\
 X^A
 &=
 L\sinh\rho\,n^A,
 &
 \sum_{A=1}^{D-1}(n^A)^2
 &=
 1 ,
\end{align}
for which the induced metric becomes
\begin{equation}
 \dd s_G^2
 =
 L^2
 \left[
 -\cosh^2\rho\,\dd\tau^2
 +\dd\rho^2
 +\sinh^2\rho\,\dd\Omega_{D-2}^2
 \right].
 \label{eq:ads-global}
\end{equation}
The point $\rho=0$ is the center of global AdS, while
$\rho\rightarrow\infty$ approaches its conformal boundary.  On the
universal covering space, the latter has topology
$\mathbb{R}\times S^{D-2}$.

For the present construction, however, it is more convenient to work in the
Poincar\'e patch.  Introducing coordinates $(z,x^i)$,
$i=0,\ldots,D-2$, the physical AdS metric takes the form
\begin{equation}
 \dd s_G^2
 =
 \frac{L^2}{z^2}
 \left(
   \dd z^2+\eta_{ij}\dd x^i\dd x^j
 \right),
 \qquad
 z>0 .
 \label{eq:adsmetric}
\end{equation}
The hypersurfaces of constant $z$ are $(D-1)$-dimensional Minkowski
slices, while $z$ parametrizes the direction into the AdS bulk.  The
conformal boundary is approached as $z\rightarrow0$.  Removing the
divergent conformal factor gives
\begin{equation}
 \lim_{z\to0}
 \frac{z^2}{L^2}\,\dd s_G^2
 =
 \eta_{ij}\dd x^i\dd x^j ,
\end{equation}
so a single Poincar\'e patch naturally selects the Minkowski representative
$\mathbb{R}^{1,D-2}$ of the boundary conformal class.

\begin{figure}[h]
    \centering
    \begin{tikzpicture}[
        scale=1.05,
        >=Latex,
        font=\small,
        boundary/.style={draw=gray!55, very thick},
        slice/.style={draw=cyan!45!blue, fill=cyan!12, opacity=0.65},
        axis/.style={->, thick, draw=gray!70},
        bulklabel/.style={font=\large}
    ]

    \coordinate (A) at (-4,0);
    \coordinate (B) at ( 4,0);
    \coordinate (C) at ( 5,1.7);
    \coordinate (D) at (-3,1.7);

    \fill[gray!8] (A)--(B)--(C)--(D)--cycle;
    \draw[boundary] (A)--(B)--(C)--(D)--cycle;

    \foreach \x in {-3,-2,...,3}{
        \draw[gray!20]
        (\x,0) -- ++(1,1.7);
    }

    \foreach \y in {0.35,0.7,1.05,1.4}{
        \draw[gray!20]
        (-4+\y/1.7,{\y}) -- (4+\y/1.7,{\y});
    }

    \node[below] at (0,0)
        {$\partial\AdS_D:\; z=0$};

    \draw[axis, very thick]
        (0.45,0.8) -- (0.45,6.1)
        node[above] {$z$};

    \node[right] at (0.55,5.45)
        {$z>0$};

    \begin{scope}[shift={(0,2.35)},scale=0.83]
        \filldraw[slice]
            (-3,0)--(3,0)--(3.7,1.15)--(-2.3,1.15)--cycle;
    \end{scope}

    \node[right] at (3.55,2.95)
        {$z=z_1$};

    \begin{scope}[shift={(0,4.0)},scale=0.62]
        \filldraw[slice]
            (-3,0)--(3,0)--(3.7,1.15)--(-2.3,1.15)--cycle;
    \end{scope}

    \node[right] at (2.55,4.45)
        {$z=z_2>z_1$};

    \draw[axis]
        (-2.6,0.43) -- (0.4,0.43)
        node[right, font=\footnotesize] {$x^i$};

    \draw[axis]
        (-2.6,0.43) -- (-2.0,1.35)
        node[above, yshift=-3pt, font=\footnotesize] {$x^j$};

    \node[bulklabel, cyan!40!blue!70!black]
        at (-2.15,5.4)
        {$\AdS_D$ bulk};

    \draw[decorate,
          decoration={brace,amplitude=6pt},
          thick,
          draw=gray!65]
        (-3.4,1.9) -- (-2.25,5.0);

    \node[rotate=70, left]
        at (-3.05,4.55)
        {$\displaystyle \frac{L^2}{z^2}$ warp factor};

    \node[right, align=left]
        at (0.7,6.2)
        {\scriptsize increasing $z$\\[-1mm]
         \scriptsize (Poincar\'e interior)};

    \end{tikzpicture}

    \caption{
    Poincar\'e patch of $\AdS_D$. The hypersurfaces of fixed radial
    coordinate $z$ are $(D-1)$-dimensional Minkowski slices parametrized
    by $x^i$. The conformal boundary lies at $z=0$, while increasing $z$
    moves towards the Poincar\'e interior.
    }
    \label{fig:poincare-ads}
\end{figure}
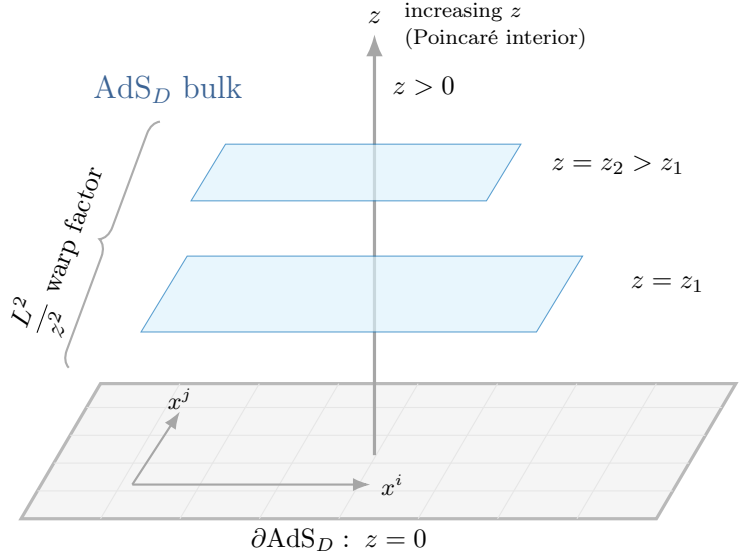

The geometry of the Poincar\'e patch is illustrated schematically in
Fig.~\ref{fig:poincare-ads}.  Constant-$z$ hypersurfaces form Minkowski
slices, the conformal boundary lies at $z=0$, and increasing $z$ moves
towards the Poincar\'e interior.

A single Poincar\'e patch does not cover the whole of global AdS.  Its
relation to the embedding coordinates can be written as
\begin{align}
 X^{-1}
 &=
 \frac{L^2+z^2+\eta_{ij}x^i x^j}{2z},
 \nonumber\\
 X^{D-1}
 &=
 \frac{L^2-z^2-\eta_{ij}x^i x^j}{2z},
 \nonumber\\
 X^i
 &=
 \frac{L}{z}\,x^i,
 \qquad
 i=0,\ldots,D-2.
 \label{eq:poincare-embedding}
\end{align}
These relations automatically satisfy the embedding constraint
\eqref{eq:ads-embedding}.  Combining them with the global parametrization
gives
\begin{equation}
 z
 =
 \frac{L}{
 \cosh\rho\,\cos\tau+\sinh\rho\,n^{D-1}
 },
 \label{eq:z-global-poincare}
\end{equation}
and therefore the patch $z>0$ corresponds to
\begin{equation}
 \cosh\rho\,\cos\tau+\sinh\rho\,n^{D-1}>0.
 \label{eq:poincare-region}
\end{equation}
The Poincar\'e coordinates therefore describe only a wedge of global AdS.
Correspondingly, the $z=0$ boundary of this patch is represented by
$\mathbb{R}^{1,D-2}$, whereas the boundary of global AdS is the cylinder
$\mathbb{R}\times S^{D-2}$.  These are related by a conformal
transformation and represent different patches of the same boundary
conformal geometry.

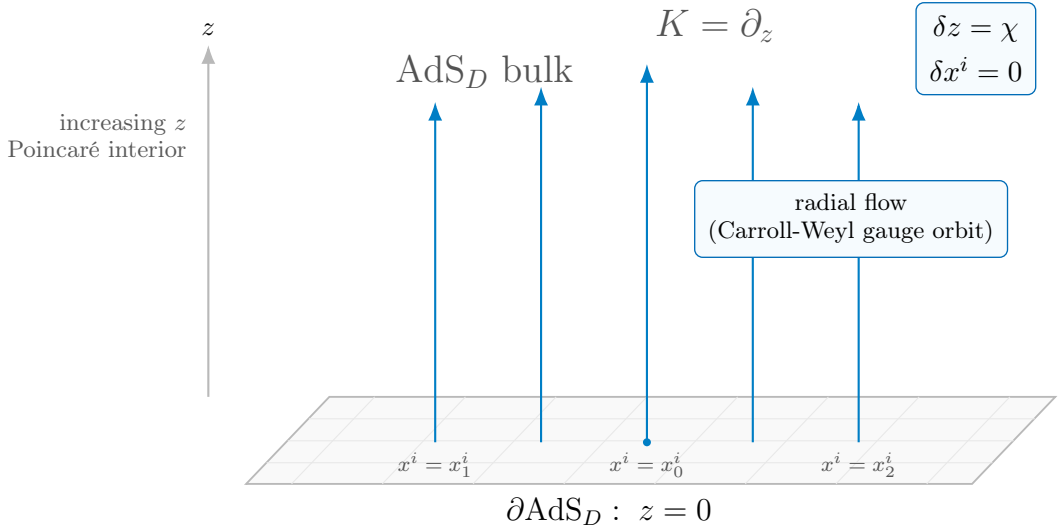
\begin{figure}[H]
\centering

\begin{tikzpicture}[
    x=1cm,
    y=1cm,
    font=\small,
    orbit/.style={
        draw=cyan!60!blue,
        line width=0.8pt,
        -{Latex[length=2.5mm]}
    },
    boundary/.style={
        draw=gray!55,
        fill=gray!5,
        line width=0.7pt
    },
    note/.style={
        draw=cyan!45!blue,
        fill=cyan!3,
        line width=0.5pt,
        rounded corners=3pt,
        inner sep=5pt,
        align=center
    },
    zaxis/.style={
        draw=gray!55,
        line width=0.8pt,
        -{Latex[length=2.5mm]}
    }
]


\coordinate (A) at (-4.8,0);
\coordinate (B) at ( 4.8,0);
\coordinate (C) at ( 5.9,1.15);
\coordinate (D) at (-3.7,1.15);

\filldraw[boundary]
    (A) -- (B) -- (C) -- (D) -- cycle;


\draw[gray!18] (-4.2,0) -- (-3.1,1.15);
\draw[gray!18] (-3.0,0) -- (-1.9,1.15);
\draw[gray!18] (-1.8,0) -- (-0.7,1.15);
\draw[gray!18] (-0.6,0) -- ( 0.5,1.15);
\draw[gray!18] ( 0.6,0) -- ( 1.7,1.15);
\draw[gray!18] ( 1.8,0) -- ( 2.9,1.15);
\draw[gray!18] ( 3.0,0) -- ( 4.1,1.15);
\draw[gray!18] ( 4.2,0) -- ( 5.3,1.15);

\draw[gray!18] (-4.53,0.28) -- (5.07,0.28);
\draw[gray!18] (-4.25,0.57) -- (5.35,0.57);
\draw[gray!18] (-3.97,0.86) -- (5.63,0.86);

\node[
    below,
    font=\large
] at (0,-0.05)
{$\partial\mathrm{AdS}_{D}:\; z=0$};


\draw[zaxis]
    (-5.3,1.15) -- (-5.3,5.8);

\node[
    above
] at (-5.3,5.8)
{$z$};

\node[
    left,
    align=right,
    text=gray!70!black
] at (-5.45,4.6)
{
    \footnotesize increasing $z$\\[-1mm]
    \footnotesize Poincar\'e interior
};


\draw[orbit]
    (-2.3,0.55) -- (-2.3,5.05);

\draw[orbit]
    (-0.9,0.55) -- (-0.9,5.25);

\draw[orbit]
    (0.5,0.55) -- (0.5,5.55);

\draw[orbit]
    (1.9,0.55) -- (1.9,5.25);

\draw[orbit]
    (3.3,0.55) -- (3.3,5.05);

\fill[cyan!60!blue]
    (0.5,0.55) circle (1.5pt);


\node[
    below,
    text=gray!65!black,
    font=\scriptsize
] at (-2.3,0.55)
{$x^i=x_1^i$};

\node[
    below,
    text=gray!65!black,
    font=\scriptsize
] at (0.5,0.55)
{$x^i=x_0^i$};

\node[
    below,
    text=gray!65!black,
    font=\scriptsize
] at (3.3,0.55)
{$x^i=x_2^i$};


\node[
    text=gray!65!black,
    font=\Large
] at (-1.65,5.45)
{$\mathrm{AdS}_{D}\ \mathrm{bulk}$};

\node[
    text=gray!65!black,
    font=\Large
] at (1.40,6.05)
{$K=\partial_z$};


\node[
    note,
    font=\normalsize
] at (4.85,5.75)
{
    $\delta z=\chi$\\[2pt]
    $\delta x^i=0$
};


\node[
    note
] at (3.20,3.50)
{
    \footnotesize radial flow\\[-1mm]
    \footnotesize (Carroll-Weyl gauge orbit)
};

\end{tikzpicture}

\caption{
Schematic picture of the radial flow in the Poincar\'e patch of
$\mathrm{AdS}_{D}$. The vector field $K=\partial_z$ generates motion
along the radial direction, while the transverse coordinates $x^i$
remain unchanged. In the Carroll-Weyl theory this radial motion is
gauged, so each radial trajectory represents a local gauge orbit,
with $\delta z=\chi$ and $\delta x^i=0$.
}
\label{fig:elegant-radial-flow}

\end{figure}

The radial structure of the Poincar\'e patch is illustrated schematically in
Fig.~\ref{fig:elegant-radial-flow}.  The integral curves of
$K=\partial_z$ move along the radial direction while leaving the transverse
coordinates $x^i$ fixed.  These curves are precisely the local
Carroll-Weyl gauge orbits, generated infinitesimally by
$\delta z=\chi$ and $\delta x^i=0$. The integral curves of $K=\partial_z$ satisfy
\begin{equation}
    \frac{dz}{d\lambda}=1,
    \qquad
    \frac{dx^i}{d\lambda}=0,
\end{equation}
and are therefore given locally by
\begin{equation}
    \gamma_{x_0}(\lambda)
    =
    \bigl(z_0+\lambda,x_0^i\bigr).
\end{equation}
Thus each orbit varies only in the radial coordinate $z$, while the
transverse coordinates $x^i$ remain fixed.

The Carroll-Weyl transformation acts pointwise along the flow of the
distinguished vector field,
\begin{equation}
    \delta_\chi X^A(\sigma)
    =
    \chi(\sigma)K^A(X(\sigma)).
\end{equation}
For $K=\partial_z$ this reduces to
\begin{equation}
    \delta_\chi z=\chi,
    \qquad
    \delta_\chi x^i=0.
\end{equation}
Hence the tangent distribution to the Carroll-Weyl gauge orbits is
$\mathrm{span}\{\partial_z\}$.  The gauge orbits therefore coincide
locally with the radial integral curves
\begin{equation}
    \Phi_\lambda(z,x^i)
    =
    (z+\lambda,x^i),
\end{equation}
and the corresponding local orbit space is parametrized by the
transverse coordinates $x^i$. For the Carroll-Weyl construction, the important feature of the
Poincar\'e description is the distinguished radial coordinate $z$.
Curves of fixed $x^i$ and varying $z$ define a natural radial flow through
the patch.  We now turn to the role of this flow in the Carroll-Weyl
gauging.

\subsection{Radial Carroll-Weyl gauging and the homothetic frame}
\label{subsec:ads-radial-gauging}

The Poincar\'e slicing (with $z>0$) naturally singles out the radial vector field
\begin{equation}
 K=\partial_z .
 \label{eq:ads-K}
\end{equation}
Its integral curves keep the boundary coordinates $x^i$ fixed and move only
along the radial direction.  With respect to the physical AdS metric,
however, this flow is not an isometry or a homothety.  Instead,
\begin{equation}
 \Lie_K G
 =
 -\frac{2}{z}G,
 \label{eq:adsckv}
\end{equation}
so that $K$ is a conformal Killing vector with
\begin{equation}
 \omega_K=-\frac{1}{z}.
 \label{eq:ads-omega}
\end{equation}
Moreover,
\begin{equation}
 G(K,K)
 =
 \frac{L^2}{z^2}>0,
 \label{eq:Knorm}
\end{equation}
and hence the selected conformal-Killing flow is spacelike throughout the
Poincar\'e patch.

The physical AdS geometry therefore belongs naturally to the
nearest-to-minimal class \cite{NullHolography}. To pass to a
minimal Carroll-Weyl realization, we introduce a conformally related metric
\begin{equation}
 g=e^{-2\Omega}G
\end{equation}
and require that $K$ become homothetic with respect to $g$.  The corresponding
frame equation is
\begin{equation}
 K(\Omega)=\omega_K-1.
 \label{eq:frameequation}
\end{equation}
For $K=\partial_z$, this gives
\begin{equation}
 \partial_z\Omega=-\frac{1}{z}-1,
\end{equation}
whose local solution is
\begin{equation}
 \Omega=-\log z-z,
 \label{eq:adsOmega}
\end{equation}
up to a function constant along the $K$-orbits.  We set this additional
function to zero.  The resulting metric is
\begin{equation}
 g
 =
 z^2e^{2z}G
 =
 L^2e^{2z}
 \left(
   \dd z^2+\eta_{ij}\dd x^i\dd x^j
 \right),
 \qquad
 \Lie_K g=2g.
 \label{eq:homframe}
\end{equation}
Thus the radial conformal-Killing flow of the physical AdS metric becomes a
homothetic flow in the conformally related frame.

Following the construction of ref.~\cite{NullHolography}, the corresponding
$K$-gauged null-string action is
\begin{equation}
 S_K
 =
 \frac{\kappa L^2}{2}
 \int_\Sigma\dd^2\sigma\,
 e^{2z}V^aV^b
 \left[
   (\partial_a z+W_a)(\partial_bz+W_b)
   +
   \eta_{ij}\partial_ax^i\partial_bx^j
 \right].
 \label{eq:Vkaction}
\end{equation}
The local Carroll-Weyl transformations are
\begin{equation}
 \delta_\chi z=\chi,
 \qquad
 \delta_\chi W_a=-\partial_a\chi,
 \qquad
 \delta_\chi V^a=-\chi V^a,
 \qquad
 \delta_\chi x^i=0.
 \label{eq:transformV}
\end{equation}
The radial coordinate $z$ therefore parametrizes the local
Carroll-Weyl orbit, while the coordinates $x^i$ are invariant under the
gauged flow.

Gauge invariance of \eqref{eq:Vkaction} is immediate.  The radial covariant
derivative satisfies
\begin{equation}
 \delta_\chi(\partial_a z+W_a)=0,
\end{equation}
while
\begin{equation}
 \delta_\chi e^{2z}=2\chi e^{2z},
 \qquad
 \delta_\chi(V^aV^b)=-2\chi V^aV^b.
\end{equation}
The variations of the two prefactors therefore cancel without using any
equation of motion.

It is convenient to introduce the gauge-invariant variables
\begin{equation}
 U^a=e^zV^a,
 \qquad
 B_a=\partial_a z+W_a.
 \label{eq:invariantvariables}
\end{equation}
They obey
\begin{equation}
 \delta_\chi U^a=0,
 \qquad
 \delta_\chi B_a=0.
\end{equation}
In terms of these variables the action becomes
\begin{equation}
 S_K
 =
 \frac{\kappa L^2}{2}
 \int_\Sigma\dd^2\sigma\,
 U^aU^b
 \left(
   B_aB_b
   +
   \eta_{ij}\partial_ax^i\partial_bx^j
 \right).
 \label{eq:invariantaction1}
\end{equation}
Thus all fields appearing in \eqref{eq:invariantaction1} are explicitly
Carroll-Weyl invariant.  The original radial coordinate appears only
through the St\"uckelberg combination $B_a$.

To expose the role of the Carroll-Weyl gauge connection, define
\begin{equation}
 b:=U^aB_a,
 \qquad
 v^i:=U^a\partial_ax^i.
 \label{eq:b-v-def}
\end{equation}
The Lagrangian density is then proportional to
\begin{equation}
 b^2+\eta_{ij}v^iv^j.
\end{equation}
Only the component of $W_a$ along the degenerate worldsheet direction
selected by $U^a$ enters the action; the component annihilated by $U^a$
drops out.

Variation with respect to $W_a$ gives
\begin{equation}
 U^aU^bB_b=0.
 \label{eq:Weom}
\end{equation}
On a worldsheet patch on which $U^a$ is nonvanishing, this implies
\begin{equation}
 U^aB_a=0.
 \label{eq:UBzero}
\end{equation}
Notice that \eqref{eq:UBzero} does not imply $B_a=0$ as a one-form.  It only
sets to zero the component of $B_a$ probed by the degenerate worldsheet
vector.

Substituting \eqref{eq:UBzero} into \eqref{eq:invariantaction1} gives the
classically reduced action
\begin{equation}
 S_{\rm red}
 =
 \frac{\kappa L^2}{2}
 \int_\Sigma\dd^2\sigma\,
 U^aU^b
 \eta_{ij}
 \partial_ax^i\partial_bx^j.
 \label{eq:ilstaction1}
\end{equation}
This is the ILST action on $(D-1)$-dimensional Minkowski space \cite{NullHolography}.  

Geometrically, the integral curves of $K=\partial_z$ keep $x^i$ fixed and
vary only the radial coordinate.  Their local orbit space is therefore
labelled by the $(D-1)$ coordinates $x^i$.  At this stage it is useful to
distinguish the orbit space from the AdS conformal boundary.  The latter
interpretation requires the additional endpoint map: every radial orbit at
fixed $x^i$ has an endpoint at $z=0$ in the conformal completion of the
physical metric $G$.  This identifies the quotient with the Minkowski
representative of the Poincar\'e conformal boundary.

For the analysis below we shall, however, retain the unreduced
Carroll-Weyl theory whenever its gauge structure, constraints and quantum
properties are relevant.  Equation \eqref{eq:ilstaction1} should therefore be
viewed as the classical reduced theory associated with the radial gauge
orbit, rather than as a replacement of the parent Carroll-Weyl
formulation.

The invariant form \eqref{eq:invariantaction1} also anticipates the
canonical description. In particular, the rescaling
$U^a=e^zV^a$ absorbs the same homothetic factor that appears in the
natural Hamiltonian constraint. This motivates the radial-adapted
Hamiltonian representative introduced in the next subsection.

\subsection{Phase-space constraints}
\label{subsec:ads-phase-space}

From the homothetic-frame metric
\eqref{eq:homframe}, the geometrically natural Hamiltonian constraint is
\begin{equation}
 \cM_g
 =
 \frac12 g^{AB}P_AP_B
 =
 \frac{e^{-2z}}{2L^2}
 \left(
   P_z^2+P_iP^i
 \right).
 \label{eq:Mg-homothetic}
\end{equation}
The explicit $z$ dependence reflects the homothetic weight of this
representative.  Introducing the smooth nonvanishing factor
\begin{equation}
 F(z)=L^2e^{2z},
\end{equation}
we define
\begin{equation}
 \cM_{\rm rad}
 :=
 F\cM_g
 =
 \frac12
 \left(
   P_z^2+P_iP^i
 \right).
 \label{eq:Mrad-definition}
\end{equation}

Since $F\neq0$ throughout the Poincar\'e patch, $\cM_g$ and
$\cM_{\rm rad}$ have the same local vanishing locus.  More generally, for
any phase-space function $A$,
\begin{equation}
 \{A,F\cM_g\}
 =
 F\{A,\cM_g\}
 +
 \{A,F\}\cM_g
 \approx
 F\{A,\cM_g\},
 \label{eq:M-rescaling-gauge-distribution}
\end{equation}
so the two constraints generate the same gauge distribution on the common
constraint surface, up to a rescaling of the corresponding gauge parameter.
Equivalently, the multiplier redefinition
\begin{equation}
 e=F^{-1}\widetilde e,
 \qquad
 \widetilde e\,\cM_g
 =
 e\,\cM_{\rm rad}
 \label{eq:M-multiplier-redefinition}
\end{equation}
implements the same change in the phase-space action.  We shall therefore
use the radial-adapted representative
\begin{equation}
 \cM\equiv\cM_{\rm rad}
\end{equation}
in the remainder of the paper.  This equivalence is local and classical;
a comparison of independently quantized representatives requires the
corresponding transformation in the extended BFV phase space.

The Carroll-Weyl constraint is fixed independently by the target-space
flow being gauged.  For a general vector field
\begin{equation}
 K=K^A(X)\partial_A,
\end{equation}
the corresponding canonical generator is
\begin{equation}
 \cS_K=K^A(X)P_A.
 \label{eq:general-K-generator}
\end{equation}
Indeed, defining
\begin{equation}
 S_K[\chi]
 =
 \frac{1}{2\pi}
 \int\dd\sigma\,
 \chi(\sigma)\,
 K^A(X)P_A,
 \label{eq:smeared-K-generator}
\end{equation}
one obtains
\begin{equation}
 \delta_\chi X^A
 =
 \{X^A,S_K[\chi]\}
 =
 \chi K^A(X),
 \label{eq:canonical-K-action}
\end{equation}
which reproduces the infinitesimal target-space flow.

For the AdS Poincar\'e construction,
\begin{equation}
 K=\partial_z,
\end{equation}
so that in the coordinates $X^A=(z,x^i)$,
\begin{equation}
 K^A=(1,0,\ldots,0).
\end{equation}
Hence
\begin{equation}
 \cS
 =
 \cS_K
 =
 P_z
 \approx0.
 \label{eq:Pzconstraint}
\end{equation}
Thus $z$ parametrizes the Carroll-Weyl gauge orbit, while the transverse
coordinates $x^i$ are invariant along the flow.

The distinction between the homothetic and radial-adapted Hamiltonian
representatives is now explicit.  For
\begin{equation}
 S[\chi]
 =
 \frac{1}{2\pi}
 \int\dd\sigma\,
 \chi P_z,
\end{equation}
the canonical brackets give
\begin{equation}
 \{S[\chi],\cM_g\}
 =
 2\chi\cM_g,
 \qquad
 \{S[\chi],\cM_{\rm rad}\}
 =
 0,
 \label{eq:SMg-SMrad}
\end{equation}
with the local delta function suppressed.  The natural homothetic
constraint is therefore covariant under the radial Carroll-Weyl flow,
whereas the radial-adapted representative is invariant.

With these constraint representatives fixed, the phase-space action of
the parent theory may be written as
\begin{equation}
 S_{\rm ph}
 =
 \frac{1}{2\pi}
 \int\dd\tau\dd\sigma\,
 \left[
   P_z\dot z
   +
   P_i\dot x^i
   -
   e\,\cM
   -
   u\,\cH
   -
   a\,\cS
 \right],
 \label{eq:phaseaction}
\end{equation}
with
\begin{align}
 \cM
 &=
 \frac12
 \left(
   P_z^2+P_iP^i
 \right),
 &
 \cH
 &=
 P_zz'+P_ix'^i,
 &
 \cS
 &=
 P_z.
 \label{eq:adsconstraints}
\end{align}
Here $\cM$ is the Hamiltonian constraint, $\cH$ generates
reparametrizations along the spatial worldsheet direction, and $\cS$
generates the local Carroll-Weyl radial flow.

The result $\cS=P_z$ is also consistent with the configuration-space
description.  There, variation of the Carroll-Weyl connection gives
\begin{equation}
 U^aB_a=0,
 \qquad
 B_a=\partial_a z+W_a.
\end{equation}
In canonical variables this condition is precisely
\begin{equation}
 P_z\approx0.
\end{equation}
An explicit verification that $P_z$ generates
$\delta_\chi z=\chi$ is given in
\cref{app:radial-constraint-checks}.

For comparison, in the Cartesian Carroll-Weyl realization in flat target
space the selected vector field is
\begin{equation}
 K=X^A\partial_A,
\end{equation}
and therefore
\begin{equation}
 \cS_{\rm flat}=X^AP_A,
\end{equation}
which generates target-space dilations.  In the present Poincar\'e-patch
realization,
\begin{equation}
 K=\partial_z,
 \qquad
 \cS_{\rm AdS}=P_z,
\end{equation}
so the Carroll-Weyl generator instead produces translations along the
radial gauge orbit.

Together with the radial-adapted Hamiltonian representative, this form of
the Carroll-Weyl generator has an immediate consequence for the
constraint algebra.  Since
\begin{equation}
 \cM
 =
 \cM_{\rm rad}
 =
 \frac12
 \left(
   P_z^2+P_iP^i
 \right)
\end{equation}
contains no explicit dependence on $z$, while $\cS=P_z$, one finds
\begin{equation}
 \{\cS,\cM\}=0.
 \label{eq:SM-local-zero}
\end{equation}
Accordingly, the Fourier modes satisfy
\begin{equation}
 [S_m,M_n]=0.
 \label{eq:noSM}
\end{equation}
This vanishing bracket is therefore a property of the radial-adapted
constraint system.  For the unrescaled homothetic representative one
instead has $\{\cS,\cM_g\}=2\cM_g$.

\begin{figure}[H]
\centering
\begin{tikzpicture}[
    >=Latex,
    line cap=round,
    line join=round,
    font=\small,
    title/.style={font=\bfseries\large}
]


\node[title] at (0,2.7)
    {\small \mdseries Punctured Euclidean space};

\draw[thin] (0,0.25) circle (0.68);
\draw[thin] (0,0.25) circle (1.12);
\draw[thin] (0,0.25) circle (1.55);
\draw[thin] (0,0.25) circle (1.85);

\fill[white] (0,0.25) circle (0.11);
\draw[semithick] (0,0.25) circle (0.11);

\draw[
    -{Latex[length=3mm]},
    semithick
]
    (0.12,0.37) -- (1.18,1.31);

\node at (0.51,0.95) {$r$};

\node at (-2.30,0.55)
      {$S^{d-1}$};

\node at (0,-1.95)
      {$\mathbb{R}^{d}\setminus\{0\}$};


\draw[
    -{Latex[length=2.2mm]},
    semithick
]
    (2.25,0.25) -- (5.15,0.25);
\node[
    draw,
    rounded corners=4pt,
    align=center,
    inner xsep=11pt,
    inner ysep=7pt,
    font=\normalsize
] at (3.70,1.25)
{
    \text{Weyl map}\\[4pt]
    $r=e^{\tau}$
};

\node[
    font=\normalsize
] at (3.70,-0.35)
{
    $\tau=\log r$
};


\begin{scope}[shift={(7.7,0.25)}]

\node[title] at (0,2.45)
    {\small \mdseries Cylinder};

\draw[semithick]
    (-1.35,-1.55) -- (-1.35,1.55);

\draw[semithick]
    (1.35,-1.55) -- (1.35,1.55);

\draw[semithick]
    (0,1.55)
    ellipse [x radius=1.35cm,y radius=0.35cm];

\draw[semithick]
    (0,-1.55)
    ellipse [x radius=1.35cm,y radius=0.35cm];

\draw[
    thin,
    dashed
]
    (0,0)
    ellipse [x radius=1.35cm,y radius=0.35cm];

\node[
    fill=white,
    inner sep=1.5pt
] at (1.08,0.08)
{
    $S^{d-1}$
};

\draw[
    -{Latex[length=3mm]},
     thin
]
    (0,-1.05) -- (0,1.05);

\node[
    right
] at (0.12,0.82)
{
    $\partial_\tau$
};

\node[
    right
] at (0.12,-0.65)
{
    $\tau$
};

\node at (0,-2.15)
      {$\mathbb{R}\times S^{d-1}$};

\end{scope}


\node[
    draw,
    rounded corners=5pt,
    inner xsep=15pt,
    inner ysep=8pt,
    font=\large
] at (3.70,-1.75)
{
    $\displaystyle
    r\partial_r=\partial_\tau
    $
};

\end{tikzpicture}

\caption{
Standard Weyl map underlying radial quantization.
Introducing $\tau=\log r$ maps
$\mathbb{R}^{d}\setminus\{0\}$ to
$\mathbb{R}\times S^{d-1}$, and the radial dilatation
generator $r\partial_r$ becomes the cylinder translation
generator $\partial_\tau$.
}
\label{fig:weyl-map-radial-cylinder}
\end{figure}
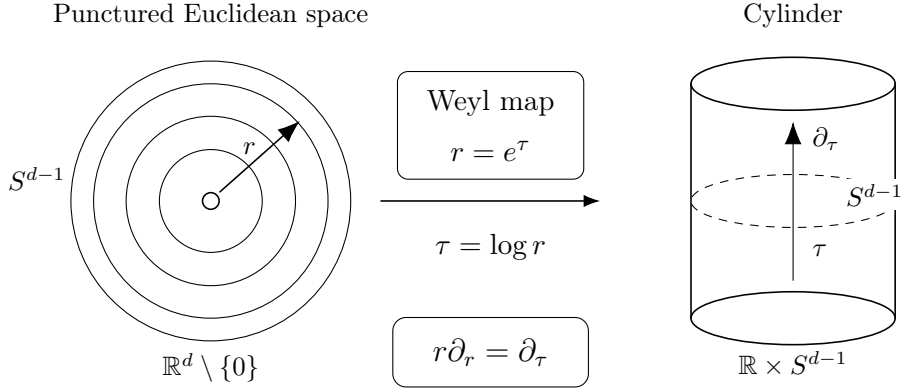

\paragraph{Analogy with radial quantization.}
A useful analogy is provided by radial quantization in conformal field theory. 
In flat Euclidean space, the dilatation generator is
\begin{equation}
    \mathcal{D} = x^\mu \partial_\mu .
\end{equation}
After the standard Weyl map
\begin{equation}
    \mathbb{R}^d\setminus\{0\}
    \longrightarrow
    \mathbb{R}\times S^{d-1},
    \qquad
    r=e^\tau ,
\end{equation}
the radial dilatation generator becomes
\begin{equation}
    r\partial_r=\partial_\tau .
\end{equation}
It is useful to recall the standard radial-quantization picture, illustrated in
Fig.~\ref{fig:weyl-map-radial-cylinder}.  The punctured Euclidean space
$\mathbb{R}^{d}\setminus\{0\}$ is foliated by spheres $S^{d-1}$ of fixed
radius $r$.  Introducing the logarithmic radial coordinate
\begin{equation}
    \tau=\log r ,
\end{equation}
or equivalently $r=e^\tau$, maps this geometry to the cylinder
$\mathbb{R}\times S^{d-1}$.  Under this map, a radial scale transformation in
flat space becomes a translation along the cylinder.  At the level of
generators,
\begin{equation}
    r\partial_r=\partial_\tau .
\end{equation}
Thus the flat-space dilatation generator is represented as the
$\tau$-translation generator in the cylindrical frame.  This provides a useful
geometric analogy for the frame-dependent interpretation of the distinguished
radial flow appearing in the Carroll-Weyl construction discussed below.

Thus a generator which appears as a dilatation in one conformal frame can be interpreted as an ordinary translation generator in another coordinate or conformal frame. This provides a useful conceptual analogy with the present construction. Here one starts from the distinguished radial flow in the AdS Poincar\'e patch and passes to a conformally related frame in which this flow becomes homothetic. The corresponding canonical generator then takes the particularly simple translation-like form
\begin{equation}
    \mathcal S=P_z .
\end{equation}
There is, however, an important distinction from the standard AdS/CFT interpretation of dilatations. The AdS isometry that induces boundary dilatations is generated by
\begin{equation}
    \mathcal{D}_{\rm AdS}
    =
    x^i\partial_i+z\partial_z ,
\end{equation}
rather than by $\partial_z$ alone. The vector field
\begin{equation}
    K=\partial_z
\end{equation}
used in the present construction is therefore not the usual AdS dilatation Killing vector. Instead, it is a particular conformal Killing vector of the Poincar\'e metric whose flow is selected for the Carroll-Weyl gauging. The analogy with radial quantization should consequently be understood at the level of the change from a dilation-like flow to a translation-like generator under a change of frame, rather than as an identification of $K=\partial_z$ with the standard AdS dilatation symmetry.

\subsection{Classical constraint algebra}
\label{subsec:ads-constraint-algebra}

We now determine the algebra generated by the three constraints
$\cM$, $\cH$ and $\cS$.  It is convenient for the subsequent OPE analysis
to introduce
\begin{equation}
 \cL=-\cH.
 \label{eq:LminusH}
\end{equation}
Using the canonical brackets
\begin{equation}
 \{z(\sigma),P_z(\sigma')\}
 =
 2\pi\delta(\sigma-\sigma'),
 \qquad
 \{x^i(\sigma),P_j(\sigma')\}
 =
 2\pi\delta^i{}_j\delta(\sigma-\sigma'),
 \label{eq:canonicalbrackets1}
\end{equation}
the corresponding Fourier modes satisfy
\begin{align}
 [L_m,L_n]
 &=
 (m-n)L_{m+n},
 &
 [L_m,M_n]
 &=
 (m-n)M_{m+n},
 \nonumber\\
 [L_m,S_n]
 &=
 -nS_{m+n},
 &
 [S_m,M_n]
 &=
 0,
 \nonumber\\
 [M_m,M_n]
 &=
 0,
 &
 [S_m,S_n]
 &=
 0.
 \label{eq:adsalgebra0}
\end{align}
The detailed derivation from the local brackets is collected in
\cref{app:ads-residual-algebra}.

The generators $(L_m,M_m)$ form the usual centerless BMS$_3$ constraint
algebra,
\begin{equation}
 [L_m,L_n]=(m-n)L_{m+n},
 \qquad
 [L_m,M_n]=(m-n)M_{m+n},
 \qquad
 [M_m,M_n]=0.
\end{equation}
The additional generators $S_n$ form an abelian current of conformal
weight one,
\begin{equation}
 [L_m,S_n]=-nS_{m+n},
 \qquad
 [S_m,S_n]=0,
\end{equation}
and commute with the BMS supertranslation sector,
\begin{equation}
 [S_m,M_n]=0.
 \label{eq:noSM}
\end{equation}

The algebra \eqref{eq:adsalgebra0} is the centerless, level-zero form of the
general BMS-Kac-Moody algebra.  The same structure appears in the
BMS-invariant free scalar theory of~\cite{Hao:2021BMS}, where an abelian
weight-one current $J_n$ satisfies
\begin{equation}
 [L_m,J_n]=-nJ_{m+n},
 \qquad
 [M_m,J_n]=0,
 \qquad
 [J_m,J_n]=0.
\end{equation}
Thus $J_n\leftrightarrow S_n$ directly reproduces
\eqref{eq:adsalgebra0}.

One-current extensions of BMS$_3$ were further classified in
ref.~\cite{Yu:2022BMSIsing}.  In their terminology, the present algebra
belongs to the Type-I class, characterized by $[M_m,J_n]=0$, whereas their
free-fermion realization gives the Type-II case with a nontrivial action of
the current on the supertranslation sector.  The same centerless algebra is
also referred to as the general BMS-Kac-Moody algebra in
ref.~\cite{Liu:2026LoopBMSKM}, whose loop construction provides a further
extension of this underlying algebra.

The mathematical structure of the mode algebra, including its interpretation as a semidirect extension of the Witt algebra and the identification of its familiar subalgebras, is summarized in~\cref{app:ads-residual-algebra}.

In the present theory, however, the current has a different geometric
origin.  The generators $S_n$ arise from the first-class Carroll-Weyl
constraint
\begin{equation}
 \cS=P_z,
\end{equation}
associated with gauging the radial flow $K=\partial_z$.  Thus the
BMS-Kac-Moody structure follows directly from the constraint algebra of
the Carroll-Weyl null string in the AdS Poincar\'e patch.

The vanishing relation \eqref{eq:noSM} is a characteristic feature of
the radial-adapted realization considered here.
This should be contrasted with the Cartesian Carroll-Weyl realization in
flat target space.  There the third constraint is
\begin{equation}
 \cS_{\rm flat}=X\cdot P,
\end{equation}
which generates dilations and acts nontrivially on the null Hamiltonian,
leading classically to
\begin{equation}
 [S_m,M_n]=2M_{m+n}
\end{equation}
up to the convention chosen for the generators.  In the present AdS
Poincar\'e realization, by contrast, $\cS=P_z$ generates a translation in
the gauge radial direction and commutes with the quadratic momentum
constraint $\cM$.

This simplification will play an important role in the BRST construction.
In particular, the radial sector $(z,P_z)$ is separated much more directly
from the transverse constraint algebra than in the Cartesian dilation
realization.  Whether this separation becomes a contractible BRST sector
at the quantum level, however, must be established from the full BRST
charge and the corresponding transformation in the extended BFV phase
space.  This quantum reduction, including the ghost sector and the role of
normal ordering, will be discussed in a companion paper \cite{Duary:2026izp}.

\subsection{Central extensions of the radial AdS residual algebra}
\label{subsec:ads-central-extensions}

It is important to distinguish the abstract residual algebra from any
particular normal-ordered quantum realization.  Before introducing its
central extensions, it is useful to characterize the centerless algebra
in terms of tensor-density modules.  Following
Refs.~\cite{OvsienkoRoger1996,Ovsienko:1998TensorDensity}, let
$\mathcal F_\lambda$ denote the space of tensor densities
$a(\sigma)(d\sigma)^\lambda$ on the circle.  The natural action of a
vector field $f(\sigma)\partial_\sigma$ is
\begin{equation}
 \mathcal L_f^{(\lambda)}a
 =
 f a' + \lambda f'a .
 \label{eq:tensordensityLie}
\end{equation}
For the mode conventions
\begin{equation}
 L_m=i e^{im\sigma}\partial_\sigma,
 \qquad
 A_n=e^{in\sigma}(d\sigma)^\lambda ,
\end{equation}
this gives
\begin{equation}
 [L_m,A_n]
 =
 -(n+\lambda m)A_{m+n}.
 \label{eq:tensordensitymodule}
\end{equation}
Comparison with
\begin{equation}
 [L_m,M_n]=(m-n)M_{m+n},
 \qquad
 [L_m,S_n]=-nS_{m+n},
 \label{eq:AdSmoduleactions}
\end{equation}
therefore shows that
\begin{equation}
 M_n\in\mathcal F_{-1},
 \qquad
 S_n\in\mathcal F_0.
 \label{eq:AdSmodules}
\end{equation}
We may thus write the centerless residual algebra as

\begin{equation}
 \mathfrak a_{\rm AdS}
 =
 \mathrm{Witt}\ltimes
 \left(
   \mathcal F_{-1}\oplus\mathcal F_0
 \right).
 \label{eq:adsalgebradef}
\end{equation}
Equivalently, in the usual conformal-weight convention, $M_n$ and
$S_n$ have weights $h_M=2$ and $h_S=1$, respectively, with
$h=1-\lambda$.

The possible central extensions are properties of the abstract Lie algebra
itself.  A Chevalley-Eilenberg analysis, modulo coboundaries, gives
\begin{equation}
 H^2(\mathfrak a_{\rm AdS};\mathbb C)
 \simeq
 \mathbb C^4.
 \label{eq:H2ads}
\end{equation}
A convenient set of representatives is
\begin{align}
 \omega_{LL}(L_m,L_n)
 &=
 \frac{1}{12}m(m^2-1)\delta_{m+n,0},
 &
 \omega_{LM}(L_m,M_n)
 &=
 \frac{1}{12}m(m^2-1)\delta_{m+n,0},
 \nonumber\\
 \omega_{LS}(L_m,S_n)
 &=
 \frac12 m(m+1)\delta_{m+n,0},
 &
 \omega_{SS}(S_m,S_n)
 &=
 m\,\delta_{m+n,0}.
 \label{eq:adsH2reps}
\end{align}
Thus the algebra admits four independent central classes, which we denote by
$\gamma_{LL}$, $\gamma_{LM}$, $\gamma_{LS}$ and $\gamma_{SS}$.

There are no independent $MM$ or $MS$ cocycles.  This already follows from
the Jacobi identities.  For example, if one assumes
\begin{equation}
 [M_m,M_n]
 =
 d(m)\delta_{m+n,0},
\end{equation}
then the $(L_k,M_m,M_n)$ Jacobi identity forces $d(m)=0$ for nonzero modes.
Likewise, a putative mixed extension
\begin{equation}
 [M_m,S_n]
 =
 e(m)\delta_{m+n,0}
\end{equation}
is ruled out by the corresponding $(L,M,S)$ Jacobi identity.  The remaining
Jacobi conditions reproduce precisely the four classes in
\eqref{eq:adsH2reps}, up to coboundaries. Three of them are
associated with the subalgebra generated by $L_n$ and $S_n$. They appear in
the $[L_m,L_n]$, $[L_m,S_n]$, and $[S_m,S_n]$ commutators, and will be
denoted by $\gamma_{LL}$, $\gamma_{LS}$, and $\gamma_{SS}$, respectively.
The remaining central term, $\gamma_{LM}$, appears in the $[L_m,M_n]$
commutator and is the usual mixed central extension of the
BMS$_3$ algebra.

The existence of the $\gamma_{LM}$ class is a characteristic feature of the
AdS radial algebra.  Here
\begin{equation}
 [S_m,M_n]=0,
\end{equation}
so the additional current does not act on the supertranslation ideal, and
the usual BMS$_3$ mixed cocycle is compatible with all Jacobi identities.
This differs from the flat Cartesian Carroll-Weyl realization, for which
\begin{equation}
 [S_m,M_n]=2M_{m+n}.
\end{equation}
In that case the $S$ generator acts nontrivially on the supertranslation
sector, and the Jacobi identities involving $(L,S,M)$ force the coefficient
of the $LM$ central term to vanish.  The $LS$ and $SS$ cocycles, however,
remain admissible.  Thus the change from the flat dilation constraint to the
AdS radial constraint enlarges the space of allowed central extensions by
restoring the standard BMS$_3$ $LM$ class.

The second cohomology \(H^2(\mathfrak a_{\rm AdS};\mathbb C)\) determines the central
extensions allowed by the abstract residual algebra, independently of any
particular quantum realization.  A specific choice of matter content,
ghost system and normal ordering instead fixes the coefficients with which
these admissible cocycles are realized.  The explicit
Chevalley-Eilenberg analysis is given in~\cref{app:adscohomology}.  In the next section we determine these
coefficients for the canonical first-order AdS realization, construct the
corresponding BRST charge, and analyze its quantum nilpotency.

\section{The radial AdS BRST complex and its quantum anomalies}
\label{sec:brst-anomaly}

We now quantize the three-constraint system generated by
$(\cL,\cM,\cS)$ in the AdS Poincar\'e realization.  We first derive the
local Faddeev-Popov ghost system associated with the three gauge
directions and then formulate the corresponding matter-ghost theory in a
highest-weight first-order representation.  The resulting OPEs determine
which of the cohomologically allowed central extensions are actually
realized and provide the quantum input for constructing the BRST charge
and testing its nilpotency.

\subsection{Radial AdS Faddeev-Popov ghosts}
\label{subsec:ads-fp}

Before constructing the operator BRST complex, we first derive the local
Faddeev-Popov ghost system of the AdS Carroll-Weyl theory.  Gauge fixing the three local
symmetries determines the corresponding ghost-antighost fields and their local couplings,
which provide the ghost sector entering the matter-ghost current algebra and the BRST
charge.

The radial Carroll-Weyl action was written in \cref{subsec:ads-radial-gauging} in terms
of the gauge-adapted variables
\begin{equation}
 U^a=e^zV^a,
 \qquad
 B_a=\partial_a z+W_a ,
 \label{eq:fp-U-B}
\end{equation}
which satisfy
\begin{equation}
 \delta_\chi U^a=0,
 \qquad
 \delta_\chi B_a=0.
 \label{eq:fp-U-B-invariant}
\end{equation}
Unlike the original vector density $V^a$, the rescaled vector density
$U^a$ is invariant under the radial Carroll-Weyl transformation. This
makes it a convenient variable for the gauge-fixing analysis.

Under an infinitesimal worldsheet diffeomorphism generated by $\epsilon^a$, $U^a$
transforms as a vector density of weight $1/2$,
\begin{equation}
 \delta_\epsilon U^a
 =
 -U^b\partial_b\epsilon^a
 +\epsilon^b\partial_bU^a
 +\frac12 U^a\partial_b\epsilon^b .
 \label{eq:fp-U-diffeo}
\end{equation}
The Carroll-Weyl connection transforms as
\begin{equation}
 \delta_\chi W_a=-\partial_a\chi ,
 \label{eq:fp-W-CW}
\end{equation}
together with its ordinary one-form transformation under worldsheet diffeomorphisms.

We choose the local gauge conditions
\begin{equation}
 G^0:=U^0-1=0,
 \qquad
 G^1:=U^1=0,
 \qquad
 G^{\rm s}:=U^aW_a=0 .
 \label{eq:fp-gauge-conditions}
\end{equation}
The first two conditions fix the worldsheet diffeomorphisms, while the last condition fixes
the radial Carroll-Weyl gauge freedom.  Notice that $U^aB_a=0$ is not a gauge condition:
since $U^a$ and $B_a$ are both Carroll-Weyl invariant, this combination is itself gauge
invariant.  Indeed, $U^aB_a=0$ arises instead from the equation of motion of the
Carroll-Weyl connection.

Evaluating \eqref{eq:fp-U-diffeo} on the gauge slice
\begin{equation}
 U^a=(1,0)
\end{equation}
gives
\begin{equation}
 \delta U^0
 =
 -\frac12\partial_\tau\epsilon^0
 +\frac12\partial_\sigma\epsilon^1,
 \qquad
 \delta U^1
 =
 -\partial_\tau\epsilon^1 .
 \label{eq:fp-U-variations}
\end{equation}
There is no Carroll-Weyl contribution to these variations because $U^a$ is invariant
under $\chi$.

For the third gauge condition, define
\begin{equation}
 W:=U^aW_a .
\end{equation}
Under diffeomorphisms this quantity transforms as a scalar density, so that its
diffeomorphism variation vanishes on the gauge slice $W=0$.  Its Carroll-Weyl variation is
\begin{equation}
 \delta_\chi W
 =
 -U^a\partial_a\chi
 \xrightarrow{\,U^a=(1,0)\,}
 -\partial_\tau\chi .
 \label{eq:fp-W-variation}
\end{equation}
Thus the three gauge functions vary as
\begin{equation}
 \begin{pmatrix}
 \delta G^0\\[1mm]
 \delta G^1\\[1mm]
 \delta G^{\rm s}
 \end{pmatrix}
 =
 \begin{pmatrix}
 -\frac12\partial_\tau&
 \frac12\partial_\sigma&
 0\\[1mm]
 0&
 -\partial_\tau&
 0\\[1mm]
 0&
 0&
 -\partial_\tau
 \end{pmatrix}
 \begin{pmatrix}
 \epsilon^0\\[1mm]
 \epsilon^1\\[1mm]
 \chi
 \end{pmatrix}.
 \label{eq:fp-gauge-variation-matrix}
\end{equation}
The local Faddeev-Popov operator is therefore
\begin{equation}
 \mathcal M_{\rm FP}^{\rm AdS}
 =
 \begin{pmatrix}
 -\frac12\partial_\tau&
 \frac12\partial_\sigma&
 0\\[1mm]
 0&
 -\partial_\tau&
 0\\[1mm]
 0&
 0&
 -\partial_\tau
 \end{pmatrix}.
 \label{eq:fp-AdS-operator}
\end{equation}

The residual transformations are the zero modes of
$\mathcal M_{\rm FP}^{\rm AdS}$ and consequently satisfy
\begin{equation}
 \partial_\tau\epsilon^1=0,
 \qquad
 \partial_\tau\chi=0,
 \qquad
 \partial_\tau\epsilon^0=\partial_\sigma\epsilon^1 .
 \label{eq:fp-residual-equations}
\end{equation}
Their general local solution is
\begin{equation}
 \epsilon^1=f(\sigma),
 \qquad
 \epsilon^0=\tau f'(\sigma)+g(\sigma),
 \qquad
 \chi=\chi(\sigma).
 \label{eq:fp-residual-solution}
\end{equation}
The three arbitrary functions $f$, $g$, and $\chi$ generate respectively the $L$, $M$,
and $S$ sectors of the residual algebra.  In particular, the radial Carroll-Weyl parameter
$\chi(\sigma)$ is independent of the supertranslation parameter $g(\sigma)$, consistently with
the classical relation
\begin{equation}
 [S_m,M_n]=0.
 \label{eq:fp-SM-zero}
\end{equation}

The Faddeev-Popov identity can be written schematically as
\begin{equation}
 1=
 \int
 \mathcal D\epsilon^0\,
 \mathcal D\epsilon^1\,
 \mathcal D\chi\,
 \delta(G^{0,\eta})
 \delta(G^{1,\eta})
 \delta(G^{{\rm s},\eta})
 \det\mathcal M_{\rm FP}^{\rm AdS},
 \label{eq:fp-identity}
\end{equation}
where $\eta=(\epsilon^0,\epsilon^1,\chi)$ denotes the combined infinitesimal gauge
transformation.

To exponentiate the determinant, introduce Grassmann-odd ghosts
\begin{equation}
 c^A=(c^0,c^1,s)
 \label{eq:fp-ghosts}
\end{equation}
and antighosts
\begin{equation}
 b_A=(2b^0,2b^1,b^{\rm s}) .
 \label{eq:fp-antighosts}
\end{equation}
The factors of two in the first two components are the standard normalization used for the
BMS ghost system \cite{Chen:2023esw}.  The Grassmann representation of the determinant gives
\begin{equation}
 \det\mathcal M_{\rm FP}^{\rm AdS}
 =
 \int
 \mathcal Db\,\mathcal Dc\,
 \exp\!\left(iS_{\rm gh}^{\rm AdS}\right),
 \label{eq:fp-det-exponentiation}
\end{equation}
with
\begin{equation}
 -b_A(\mathcal M_{\rm FP}^{\rm AdS})^{A}{}_{B}c^B
 =
 b^0\partial_\tau c^0
 -b^0\partial_\sigma c^1
 +2b^1\partial_\tau c^1
 +b^{\rm s}\partial_\tau s .
 \label{eq:fp-pre-ibp}
\end{equation}
After integrating by parts on the closed-string cylinder, the ghost action becomes
\begin{equation}
 S_{\rm gh}^{\rm AdS}
 =
 \frac{i}{2\pi}
 \int d\tau d\sigma\,
 \left(
 c^0\partial_\tau b^0
 -c^1\partial_\sigma b^0
 +2c^1\partial_\tau b^1
 +s\partial_\tau b^{\rm s}
 \right).
 \label{eq:fp-AdS-ghost-action}
\end{equation}

The first three terms in \eqref{eq:fp-AdS-ghost-action} form the usual BMS
$b c$ system, while the last term is the kinetic term for the radial Carroll-Weyl ghost.
A characteristic feature of the AdS radial realization is the absence of an algebraic
mixing term between $s$ and $b^0$.  This follows directly from the block-triangular
structure of \eqref{eq:fp-AdS-operator}: the gauge conditions on $U^a$ are not shifted by
the radial Carroll-Weyl transformation because $U^a$ is Carroll-Weyl invariant.

This should be contrasted with the Cartesian flat-space realization.  There the temporal
gauge is imposed directly on $V^a$, which transforms nontrivially under Carroll-Weyl
scaling.  The corresponding Faddeev-Popov operator therefore contains an additional
off-diagonal entry and the ghost action contains an algebraic mixing between the
Carroll-Weyl ghost and the temporal BMS antighost.  Its absence in
\eqref{eq:fp-AdS-ghost-action} is the path-integral counterpart of the simpler radial
constraint algebra found here.

On a fixed Carrollian-time slice the ghost fields are identified with the first-order
systems used below according to
\begin{equation}
 c^1\leftrightarrow c,
 \qquad
 c^0\leftrightarrow\widetilde c,
 \qquad
 s\leftrightarrow\gamma,
 \label{eq:fp-ghost-identification}
\end{equation}
and
\begin{equation}
 b^1\leftrightarrow b,
 \qquad
 b^0\leftrightarrow\widetilde b,
 \qquad
 b^{\rm s}\leftrightarrow\beta .
 \label{eq:fp-antighost-identification}
\end{equation}
The corresponding weights are therefore
\begin{equation}
 (b,c):(2,-1),
 \qquad
 (\widetilde b,\widetilde c):(2,-1),
 \qquad
 (\beta,\gamma):(1,0),
 \label{eq:fp-ghost-weights}
\end{equation}
in agreement with the BRST realization of the residual algebra developed in the next
subsection.

The calculation above determines the local Faddeev-Popov determinant in
the variables $U^a=e^zV^a$ and thereby fixes the ghost content entering
the BRST analysis below. Since the change from $V^a$ to $U^a$ is
field dependent, its implications for the full functional measure are
naturally part of the corresponding quantum measure analysis.

\subsection{Matter and ghost currents}
\label{subsec:ads-matter-ghost-currents}

We begin by specifying the quantum realization of the classical AdS
constraint algebra.  As in the flat Carroll-Weyl theory, the operator
products are defined on a fixed Carrollian-time slice.  Introducing the
formal Laurent coordinate
\begin{equation}
 w=e^{i\sigma},
\end{equation}
the Fourier modes on the spatial circle are collected into Laurent fields of
definite Witt weight.  The variable $w$ is therefore an auxiliary
mode-space coordinate and does not imply a relativistic holomorphic
factorization of the Carrollian worldsheet.  We adopt the highest-weight
first-order representation and use the same normal-ordering prescription
for the matter fields, ghosts, and BRST current
\cite{BagchiThreeVacua2020,CasaliTourkine2016,ChenHuAnomalies,DuaryMajiBRST}.

For notational convenience, in what follows we write
\begin{equation}
 u=w-v.
 \label{eq:ads-u}
\end{equation}

\paragraph{Matter fields and currents.}

Let
\begin{equation}
 X^A=(z,x^i),
 \qquad
 P_A=(P_z,P_i),
 \qquad
 A=0,\ldots,D-1 ,
\end{equation}
with elementary first-order products
\begin{equation}
 P_A(w)X^B(v)
 \sim
 -\frac{\delta_A{}^B}{u},
 \qquad
 X^B(w)P_A(v)
 \sim
 \frac{\delta_A{}^B}{u},
 \label{eq:matterope}
\end{equation}
while $X^A(w)X^B(v)$ and $P_A(w)P_B(v)$ are regular.  Differentiating
\eqref{eq:matterope} gives
\begin{equation}
 P_A(w)\partial X^B(v)
 \sim
 -\frac{\delta_A{}^B}{u^2},
 \qquad
 \partial X^B(w)P_A(v)
 \sim
 -\frac{\delta_A{}^B}{u^2}.
 \label{eq:ads-differentiated-matter-ope}
\end{equation}

The three classical constraints are promoted by point splitting to
\begin{equation}
 T^X=-:P_A\partial X^A:,
 \qquad
 M^X=\frac12:P_AP^A:,
 \qquad
 S^X=P_z .
 \label{eq:mattercurrents}
\end{equation}
These are the matter currents associated respectively with the
$\cL$, $\cM$, and $\cS$ constraints, with $S^X=P_z$ representing the
Carroll-Weyl radial constraint.

Before computing the current algebra, it is useful to verify the action of
$T^X$ on the elementary fields:
\begin{align}
 T^X(w)X^A(v)
 &\sim
 \frac{\partial X^A(v)}{u},
 \nonumber\\
 T^X(w)P_A(v)
 &\sim
 \frac{P_A(v)}{u^2}
 +\frac{\partial P_A(v)}{u}.
 \label{eq:ads-matter-Ward}
\end{align}
Hence $X^A$ and $P_A$ have Witt weights zero and one, respectively.  It
follows that $M^X$ has weight two and $S^X$ has weight one.

The action of the radial current is qualitatively different from that of the
flat Cartesian Carroll-Weyl current.  From $S^X=P_z$,
\begin{equation}
 S^X(w)z(v)
 \sim
 -\frac1u,
 \qquad
 S^X(w)x^i(v)\sim0,
 \qquad
 S^X(w)P_A(v)\sim0.
 \label{eq:ads-S-elementary}
\end{equation}
Thus $S^X$ generates translations along the radial gauge orbit rather
than a target-space dilation. In particular, the momenta carry no
$S$-charge, which already anticipates the vanishing of the $SM$ bracket.

\paragraph{Matter current algebra.}

We now evaluate the OPEs among
$(T^X,M^X,S^X)$.  Single contractions reproduce the classical residual
algebra, while possible field-independent higher-order poles arise from
double contractions.

The stress-tensor product is
\begin{equation}
 T^X(w)T^X(v)
 \sim
 \frac{D}{u^4}
 +\frac{2T^X(v)}{u^2}
 +\frac{\partial T^X(v)}{u}.
 \label{eq:TTmatter}
\end{equation}
The mixed product with the null-Hamiltonian current is
\begin{equation}
 T^X(w)M^X(v)
 \sim
 \frac{2M^X(v)}{u^2}
 +\frac{\partial M^X(v)}{u}.
 \label{eq:TMmatter}
\end{equation}
Likewise,
\begin{equation}
 T^X(w)S^X(v)
 \sim
 \frac{S^X(v)}{u^2}
 +\frac{\partial S^X(v)}{u}.
 \label{eq:TSmatter}
\end{equation}
The remaining three products are regular,
\begin{align}
 M^X(w)M^X(v)&\sim0,
 \label{eq:MMmatter}\\
 S^X(w)M^X(v)&\sim0,
 \label{eq:SMmatter}\\
 S^X(w)S^X(v)&\sim0.
 \label{eq:SSmatter}
\end{align}

The pattern of central terms is especially transparent in the first-order
realization.  The double contraction in $T^XT^X$ gives $D/u^4$.  By
contrast, $M^X$ contains only momenta, so a double contraction in
$T^XM^X$ is impossible because $P_AP_B$ is regular.  Furthermore,
$S^X=P_z$ is linear in the elementary fields, excluding the double
contractions required for either an $LS$ or an $SS$ central pole.

Using the convention in which
\begin{equation}
 T(w)T(v)
 \sim
 \frac{c_{LL}/2}{u^4}+\cdots,
\end{equation}
the matter anomaly coefficients are therefore
\begin{equation}
 (c_{LL},c_{LM},c_{LS},c_{SS})_X
 =
 (2D,0,0,0).
 \label{eq:adsmatteranomalies}
\end{equation}
The vanishing of the last three entries is a property of this particular
first-order realization; the corresponding cocycles remain allowed by the
abstract algebra discussed in the previous section.

\paragraph{Ghost fields and stress tensor.}

Each of the three bosonic first-class constraints requires a fermionic
ghost-antighost pair.  We introduce
\begin{equation}
 (b,c),
 \qquad
 (\widetilde b,\widetilde c),
 \qquad
 (\beta,\gamma),
 \label{eq:ghostpairs}
\end{equation}
associated respectively with $(L,M,S)$.  Their conformal weights are
\begin{equation}
 (b,c):(2,-1),
 \qquad
 (\widetilde b,\widetilde c):(2,-1),
 \qquad
 (\beta,\gamma):(1,0).
 \label{eq:ghostweights}
\end{equation}
All three systems are Grassmann odd; in particular, the notation
$(\beta,\gamma)$ should not be confused with the bosonic superconformal
$\beta\gamma$ system.

The elementary contractions are
\begin{equation}
 b(w)c(v)\sim\frac1u,
 \qquad
 \widetilde b(w)\widetilde c(v)\sim\frac1u,
 \qquad
 \beta(w)\gamma(v)\sim\frac1u,
 \label{eq:ghost-opes}
\end{equation}
with all cross-species products regular.

For a fermionic first-order pair $(B,C)$ of weights $(h,1-h)$, the
stress tensor fixed by the corresponding Ward identities is
\begin{equation}
 T_h
 =
 -h:B\partial C:
 -(h-1):(\partial B)C:,
 \label{eq:generic-ghost-stress}
\end{equation}
and its central charge is
\begin{equation}
 c_{\rm gh}(h)
 =
 -2(6h^2-6h+1).
 \label{eq:generic-ghost-central}
\end{equation}
The diagonal ghost stress tensor is therefore
\begin{align}
 T^{\rm gh}
 ={}&
 -2:b\partial c:
 -:(\partial b)c:
 -2:\widetilde b\partial\widetilde c:
 -:(\partial\widetilde b)\widetilde c:
 \nonumber\\
 &-:\beta\partial\gamma: .
 \label{eq:Tghost}
\end{align}
The three contributions to its central charge are
\begin{equation}
 c_{bc}=-26,
 \qquad
 c_{\widetilde b\widetilde c}=-26,
 \qquad
 c_{\beta\gamma}=-2,
\end{equation}
and hence
\begin{equation}
 c_{LL}^{\rm gh}=-54.
 \label{eq:ghostcentral}
\end{equation}

\paragraph{Semidirect ghost currents.}

The complete ghost algebra is not obtained merely by summing three
independent first-order systems.  The currents associated with $M$ and $S$
are fixed by the structure constants of the classical algebra.  The BRST
variations of the ghost coordinates take the form
\begin{align}
 \delta_B c
 &=
 c\partial c,
 \nonumber\\
 \delta_B\widetilde c
 &=
 c\partial\widetilde c
 +\widetilde c\,\partial c,
 \nonumber\\
 \delta_B\gamma
 &=
 c\partial\gamma .
 \label{eq:ads-ghost-transformations}
\end{align}
In particular, there is no term proportional to
$\gamma\widetilde c$ in $\delta_B\widetilde c$, reflecting the absence of an
$SM$ structure constant.

A convenient local representative of the corresponding ghost currents is
\begin{equation}
 S^{\rm gh}
 =
 \partial:c\beta:,
 \qquad
 M^{\rm gh}
 =
 :c\partial\widetilde b:
 +2:(\partial c)\widetilde b: .
 \label{eq:otherghostcurrents}
\end{equation}
These expressions differ from those of the flat dilation complex. In the
flat theory, the nonvanishing bracket $[S_m,M_n]=2M_{m+n}$ produces
additional terms involving the $M$-ghost sector. Their absence here
reflects the vanishing $SM$ bracket of the present residual algebra.

\paragraph{Ghost current algebra.}

The independent ghost OPEs are
\begin{equation}
 T^{\rm gh}(w)T^{\rm gh}(v)
 \sim
 -\frac{27}{u^4}
 +\frac{2T^{\rm gh}(v)}{u^2}
 +\frac{\partial T^{\rm gh}(v)}{u},
 \label{eq:TTghost}
\end{equation}
\begin{equation}
 T^{\rm gh}(w)M^{\rm gh}(v)
 \sim
 \frac{2M^{\rm gh}(v)}{u^2}
 +\frac{\partial M^{\rm gh}(v)}{u},
 \label{eq:TMghost}
\end{equation}
and
\begin{equation}
 T^{\rm gh}(w)S^{\rm gh}(v)
 \sim
 \frac{S^{\rm gh}(v)}{u^2}
 +\frac{\partial S^{\rm gh}(v)}{u}.
 \label{eq:TSghost}
\end{equation}
The remaining products are regular,
\begin{align}
 M^{\rm gh}(w)M^{\rm gh}(v)&\sim0,
 \label{eq:MMghost}\\
 S^{\rm gh}(w)M^{\rm gh}(v)&\sim0,
 \label{eq:SMghost}\\
 S^{\rm gh}(w)S^{\rm gh}(v)&\sim0.
 \label{eq:SSghost}
\end{align}

The absence of the additional central poles follows directly from the field
content of the currents.  The current
$S^{\rm gh}=\partial:c\beta:$ involves fields from the independent
$(b,c)$ and $(\beta,\gamma)$ pairs.  It therefore has no self-contraction,
and no double contraction with a single term in $T^{\rm gh}$ is possible.
Similarly, $M^{\rm gh}$ contains $c$ and $\widetilde b$, but not both
members of either canonical pair, so neither $M^{\rm gh}M^{\rm gh}$ nor the
mixed $T^{\rm gh}M^{\rm gh}$ product develops an additional central pole.
Since $S^{\rm gh}$ and $M^{\rm gh}$ share no canonically conjugate fields,
their mutual OPE is also regular.

The ghost anomaly coefficients are consequently
\begin{equation}
 (c_{LL},c_{LM},c_{LS},c_{SS})_{\rm gh}
 =
 (-54,0,0,0).
 \label{eq:adsghostanomalies}
\end{equation}

\paragraph{Total current algebra.}

Matter and ghost fields are independent, so their central terms add. We define
\begin{equation}
 T^{\rm tot}=T^X+T^{\rm gh},
 \qquad
 M^{\rm tot}=M^X+M^{\rm gh},
 \qquad
 S^{\rm tot}=S^X+S^{\rm gh}.
 \label{eq:ads-total-currents}
\end{equation}
The total OPEs take the same non-central form as the classical residual
algebra,
\begin{align}
 T^{\rm tot}(w)T^{\rm tot}(v)
 &\sim
 \frac{2D-54}{2u^4}
 +\frac{2T^{\rm tot}(v)}{u^2}
 +\frac{\partial T^{\rm tot}(v)}{u},
 \nonumber\\
 T^{\rm tot}(w)M^{\rm tot}(v)
 &\sim
 \frac{2M^{\rm tot}(v)}{u^2}
 +\frac{\partial M^{\rm tot}(v)}{u},
 \nonumber\\
 T^{\rm tot}(w)S^{\rm tot}(v)
 &\sim
 \frac{S^{\rm tot}(v)}{u^2}
 +\frac{\partial S^{\rm tot}(v)}{u},
 \nonumber\\
 M^{\rm tot}(w)M^{\rm tot}(v)
 &\sim0,
 \qquad
 S^{\rm tot}(w)M^{\rm tot}(v)\sim0,
 \qquad
 S^{\rm tot}(w)S^{\rm tot}(v)\sim0.
 \label{eq:ads-total-current-algebra}
\end{align}
Thus the anomaly vector realized by the canonical first-order AdS complex is
\begin{equation}
 (c_{LL},c_{LM},c_{LS},c_{SS})_{\rm tot}
 =
 (2D-54,0,0,0).
 \label{eq:totaladsanomaly}
\end{equation}

The abstract residual algebra admits four independent central classes, but
the matter-ghost representation considered here realizes only the $LL$
cocycle. We now determine how this remaining central term enters the
BRST square and the resulting condition for cancellation of the
nontrivial quantum obstruction.

\subsection{BRST current and nilpotency}
\label{subsec:ads-brst-nilpotency}

Having determined the complete matter-ghost current algebra, we can now
construct the BRST operator and test its quantum nilpotency.  The classical
part of the construction is fixed entirely by the centerless constraint
algebra, while the possible obstruction to nilpotency is controlled by the
central terms determined above.

\paragraph{BRST current.}

For a first-class algebra with generators $G_\alpha$ and structure
constants $f_{\alpha\beta}{}^\delta$, the minimal BFV charge takes the
mode-space form
\begin{equation}
 Q_{\rm cl}
 =
 C^\alpha G_\alpha
 -\frac12
 C^\alpha C^\beta
 f_{\alpha\beta}{}^\delta B_\delta,
 \qquad
 \{G_\alpha,G_\beta\}
 =
 f_{\alpha\beta}{}^\delta G_\delta ,
 \label{eq:bfv-mode-formula}
\end{equation}
where the condensed index includes both the type of constraint and its
Fourier mode.

Applying this construction to the AdS algebra generated by $(L,M,S)$ gives
the local current
\begin{align}
 j_B={}&
 :cT^X:
 +:\widetilde cM^X:
 +:\gamma S^X:
 +:bc\partial c:
 +:\beta c\partial\gamma:
 \nonumber\\
 &\quad
 +:\widetilde b c\partial\widetilde c:
 +:\widetilde b\widetilde c\partial c:,
 \label{eq:brstcurrent}
\end{align}
and
\begin{equation}
 Q_B
 =
 \oint\frac{\dd w}{2\pi i}\,j_B(w).
 \label{eq:ads-brst-charge}
\end{equation}

The cubic ghost terms are fixed by the nonvanishing brackets of the
centerless constraint algebra.  In particular, the $[L,L]$ bracket gives
the usual $:bc\partial c:$ interaction, the $[L,M]$ bracket gives
$
:\widetilde b\,c\,\partial\widetilde c:
+:\widetilde b\,\widetilde c\,\partial c:
$,
and the $[L,S]$ bracket gives the term
$
:\beta\,c\,\partial\gamma:
$.
By contrast,
\begin{equation}
 [S_m,M_n]=0,
\end{equation}
so there is no structure constant mixing the $S$ and $M$ sectors and hence
no corresponding cubic ghost vertex.  In particular, the term
$2:\gamma\widetilde b\widetilde c:$ that appears in the flat
Carroll-Weyl dilation complex, where $[S_m,M_n]=2M_{m+n}$, is absent in
the AdS radial realization\footnote{For the unrescaled homothetic constraint $\cM_g$, the $\cS$--$\cM_g$ bracket is nonvanishing, with a corresponding ghost coupling. Relating the two quantum formulations requires the associated field-dependent transformation in the extended BFV phase space.}

As usual, the local representative of $j_B$ is defined only up to total
derivatives of ghost-number-one operators.  Such improvements can modify
local OPEs but do not change the contour charge $Q_B$.

\paragraph{Antighost identities.}

A useful check of the construction is provided by the OPE of $j_B$ with the
three antighosts.  Taking the contour residues gives
\begin{equation}
 \{Q_B,b\}
 =
 T^{\rm tot},
 \qquad
 \{Q_B,\widetilde b\}
 =
 M^{\rm tot},
 \qquad
 \{Q_B,\beta\}
 =
 S^{\rm tot}.
 \label{eq:ads-antighost-identities}
\end{equation}
Thus the three total currents obtained in the previous subsection are
precisely the BRST variations of the corresponding antighosts.  This is the
three-constraint analogue of the familiar relation
$\{Q_B,b\}=T^{\rm tot}$ in the tensile bosonic string.

At the classical level the charge is nilpotent.  Indeed, the
constraint-valued contribution to $Q_B^2$ is cancelled by the cubic ghost
terms, while the remaining three-ghost contribution is proportional to the
Jacobi identity of the centerless algebra.  Quantum mechanically, this
cancellation can fail when the total constraint algebra acquires a central
extension.

\paragraph{Quantum nilpotency.}

The previous subsection showed that the canonical first-order AdS
realization has
\begin{equation}
 (c_{LL},c_{LM},c_{LS},c_{SS})_{\rm tot}
 =
 (2D-54,0,0,0),
\end{equation}
as given in eq.~\eqref{eq:totaladsanomaly}.  More generally, each
nontrivial central cocycle realized by a quantum representation produces an
independent ghost-number-two contribution to $Q_B^2$ and must therefore
vanish for the BRST charge to define a differential.

For the canonical realization considered here only the Virasoro-type
cocycle survives.  Its contribution to the square of the BRST charge is
\begin{equation}
 Q_B^2\big|_{\rm cocycle}
 =
 \frac{2D-54}{24}
 \sum_{m\in\mathbb Z}
 m(m^2-1)c_{-m}c_m,
 \label{eq:Qsquare}
\end{equation}
up to the common normalization associated with the mode conventions.
Since the operators $c_{-m}c_m$ are nonvanishing independent elements of
the ghost algebra, cancellation of this obstruction requires
\begin{equation}
 2D-54=0,
\end{equation}
and hence
\begin{equation}
 D=27.
 \label{eq:D27}
\end{equation}

This condition cancels the nontrivial $LL$ cocycle in the present
representation.  It should be distinguished from possible zero-mode
normal-ordering conditions.  Shifts of constraint zero modes modify the
coboundary part of the mode algebra and can generate terms linear in the
mode number, but they cannot cancel the cubic
$m(m^2-1)$ Virasoro cocycle.  The zero-mode condition must therefore be
fixed separately for complete BRST nilpotency.

The contrast with the flat Carroll-Weyl complex is now manifest. In the
present realization, the radial current $S^X=P_z$ produces no matter
$LS$ or $SS$ anomaly, while the vanishing $SM$ bracket removes the
corresponding additional ghost couplings. Consequently, in the canonical
highest-weight representation considered here, the nontrivial quantum
obstruction reduces to the single $LL$ channel.

\subsection{Comparison with the minimal flat Carroll-Weyl result}
\label{subsec:flat-comparison}

The value $D=27$ appearing in the AdS radial BRST complex is numerically
the same as the value at which the Virasoro-type anomaly vanishes in the
flat Carroll-Weyl theory.  The two results, however, have different
meanings.  In the flat theory $D=27$ cancels only one component of a
three-channel anomaly vector, whereas in the canonical AdS radial
realization the non-Virasoro anomaly coefficients vanish identically.
It is therefore useful to compare the two BRST complexes more closely and
identify the structural origin of this difference.

\paragraph{Different third constraints.}

In the flat minimal Carroll-Weyl theory the third current is the target-space
dilation current,
\begin{equation}
 S_{\rm flat}^X=:X^A P_A:,
\end{equation}
whereas in the AdS Poincar\'e realization considered here it is the radial
translation current,
\begin{equation}
 S_{\rm AdS}^X=P_z.
\end{equation}
Thus
\begin{equation}
 S_{\rm flat}^X=:X^A P_A:,
 \qquad
 S_{\rm AdS}^X=P_z.
 \label{eq:twogenerators}
\end{equation}
The distinction is already visible at the level of elementary OPEs.
The flat current acts nontrivially on every first-order pair $(X^A,P_A)$,
whereas $P_z$ translates only the radial coordinate and acts trivially on
the momenta.  Consequently,
\begin{equation}
 S_{\rm flat}^X(w)S_{\rm flat}^X(v)
 \sim
 -\frac{D}{(w-v)^2},
 \qquad
 S_{\rm AdS}^X(w)S_{\rm AdS}^X(v)
 \sim0.
 \label{eq:twoss}
\end{equation}

An equally important difference is the action on the null-Hamiltonian
current.  In the flat theory,
\begin{equation}
 [S_m,M_n]=2M_{m+n},
 \label{eq:flatSMcomparison}
\end{equation}
while in the AdS radial algebra,
\begin{equation}
 [S_m,M_n]=0.
 \label{eq:adsSMcomparison}
\end{equation}
The flat bracket \eqref{eq:flatSMcomparison} generates the cubic ghost
interaction
$2:\gamma\widetilde b\widetilde c:$ and the corresponding additional
terms in the ghost currents.  These terms are absent in the AdS radial
complex because there is no $SM$ structure constant.

\paragraph{Comparison of the anomaly coefficients.}

These differences in the matter current and in the residual constraint
algebra propagate directly into the anomaly structure. In the highest-weight first-order representation, the flat
matter currents give
\begin{equation}
 (c_{LL},c_{LS},c_{SS})_{\rm flat}^{X}
 =
 (2D,-D,-D),
\end{equation}
whereas the complete flat ghost system contributes
\begin{equation}
 (c_{LL},c_{LS},c_{SS})_{\rm flat}^{\rm gh}
 =
 (-54,6,4).
\end{equation}
Hence
\begin{equation}
 (c_{LL},c_{LS},c_{SS})_{\rm flat}^{\rm tot}
 =
 (2D-54,\,6-D,\,4-D).
 \label{eq:flat-total-anomaly-comparison}
\end{equation}
The three independent conditions would require
\begin{equation}
 D=27,
 \qquad
 D=6,
 \qquad
 D=4,
\end{equation}
respectively, and therefore have no common solution.

By contrast, the canonical AdS radial realization gives
\begin{equation}
 (c_{LL},c_{LM},c_{LS},c_{SS})_{\rm AdS}^{X}
 =
 (2D,0,0,0),
\end{equation}
and
\begin{equation}
 (c_{LL},c_{LM},c_{LS},c_{SS})_{\rm AdS}^{\rm gh}
 =
 (-54,0,0,0).
\end{equation}
Thus
\begin{equation}
 (c_{LL},c_{LM},c_{LS},c_{SS})_{\rm AdS}^{\rm tot}
 =
 (2D-54,0,0,0),
 \label{eq:ads-total-anomaly-comparison}
\end{equation}
and the nontrivial cocycle obstruction vanishes at $D=27$. The comparison is summarized in table~\ref{tab:comparison}.

\begin{table}[H]
\centering
\small
\renewcommand{\arraystretch}{1.25}
\begin{tabular}{
|>{\raggedright\arraybackslash}p{3.3cm}
|p{2.0cm}
|p{1.7cm}
|p{2.65cm}
|p{2.25cm}|}
\hline
\text{Constraint realization}
&
\text{Third current}
&
\text{$[S,M]$}
&
\text{Matter anomaly}
&
\text{Ghost anomaly}
\\
\hline

Minimal flat Carroll-Weyl string
&
$X\!\cdot P$
&
$2M$
&
$(2D,-D,-D)$
&
$(-54,6,4)$
\\
\hline

AdS Poincar\'e radial complex
&
$P_z$
&
$0$
&
$(2D,0,0,0)$
&
$(-54,0,0,0)$
\\
\hline


\end{tabular}
\caption{
Anomaly coefficients in the same highest-weight first-order convention.
For the flat theory the entries are ordered as $(LL,LS,SS)$, while for the
AdS radial theory they are ordered as $(LL,LM,LS,SS)$. The flat and radial AdS
rows describe different BRST complexes.
}
\label{tab:comparison}
\end{table}

In particular, at $D=27$ the flat anomaly vector is
\begin{equation}
 (c_{LL},c_{LS},c_{SS})_{\rm flat}^{\rm tot}\big|_{D=27}
 =
 (0,-21,-23).
 \label{eq:flatD27}
\end{equation}
Thus $D=27$ cancels only the Virasoro-type obstruction in the flat
Carroll-Weyl complex; the mixed $LS$ and affine $SS$ anomalies remain.
The value $D=27$ therefore has a different status in the two theories.
In the AdS radial realization it cancels the only cocycle realized by the
canonical matter-ghost representation, whereas in the flat dilation
complex it cancels only one of three independent obstructions.

\section{Conclusions and future directions}
\label{sec:conclusions}

We have studied the Carroll-Weyl gauged null string in the Poincar\'e
patch of $\AdS_D$, focusing on the residual constraint algebra and its
BRST quantization.  The radial conformal Killing vector
$K=\partial_z$ can be brought locally to a homothetic form by passing to
a suitable conformally related metric.  In the resulting Carroll-Weyl
description the third first-class constraint is $\cS=P_z$,
so that it generates translations along the radial gauge orbit rather than the target-space dilations generated by $X\cdot P$ in the Cartesian flat-space realization.

With this radial realization, the generators satisfy $[S_m,M_n]=0$, and the residual symmetry algebra becomes a $\mathrm{BMS}_3$ algebra extended by an abelian weight-one current.  At the abstract
level the algebra admits four independent central classes, in the
$LL$, $LM$, $LS$, and $SS$ channels.  This illustrates the distinction
between the central extensions allowed by the residual algebra and the
particular levels realized after a quantum representation has been
chosen.

For the highest-weight first-order representation considered here, the
matter and ghost sectors combine to give $
 (c_{LL},c_{LM},c_{LS},c_{SS})_{\rm tot}
 =
 (2D-54,0,0,0)$. Thus only the Virasoro-type cocycle is realized, and its cancellation
selects $D=27$. This differs qualitatively from the
minimal flat Carroll-Weyl complex, where the dilation current
$X\cdot P$ generates additional $LS$ and $SS$ anomaly channels.
The difference can therefore be traced to the different realization
of the Carroll-Weyl generator and the corresponding change in the
ghost couplings, rather than simply to the presence of an additional
constraint.

A natural next step is to complete the quantum BRST analysis by fixing
the zero-mode and normal-ordering conditions and determining the resulting
BRST cohomology and physical state space. It would also be useful to
establish explicitly the quantum BFV relation to the unrescaled
homothetic formulation, including the accompanying ghost transformation
and functional measure.

A second question concerns the relation between the unreduced
Carroll-Weyl theory and its gauge-reduced ILST description.  At the
classical level the radial direction can be eliminated explicitly, but
a quantum comparison also requires control of the induced functional
measure, regulator, normal ordering, and ghost sector.  The relation to
the $(D-1)$-dimensional reduced theory, including the $27\rightarrow26$ correspondence, is studied in a companion work \cite{Duary:2026izp}.

The present construction is also intrinsically local.  The conformal
frame in which the radial flow becomes homothetic is defined patchwise,
and a single Poincar\'e patch covers only part of global AdS.  Extending
the analysis to global worldsheet configurations and clarifying the role
of radial zero modes and boundary conditions are therefore important
steps toward a genuinely global formulation.  Ultimately, the
parent-reduced correspondence should be tested at the level of
physical observables, where possible differences in the measure and
global data can no longer be ignored.

Finally, the curved-background construction is not restricted to the
specific radial flow used here.  More general conformal Killing vectors
define different Carroll-Weyl gauge orbits and may lead to distinct
residual algebras and quantum anomaly structures.  Exploring such
realizations, as well as other backgrounds admitting suitable
conformal-Killing flows, should help determine which aspects of the
present AdS construction are universal and which are specific to the
radial Poincar\'e realization.

\section*{Acknowledgments}
We thank M.~M.~Sheikh-Jabbari for his valuable comments on our draft. SD is supported by the Shuimu Tsinghua Scholar Program of Tsinghua University and the Beijing Natural Science Foundation of China under Grant No.~IS25035.

\appendix

\section{Killing, homothetic and conformal Killing vector fields}
\label{app:Killing-homothety-CKV}

In this appendix we collect the geometric facts about Killing, homothetic
and conformal Killing vector fields that are relevant for the
Carroll-Weyl-gauged null-string construction.

Let $(\mathcal{M},g)$ be a $D$-dimensional pseudo-Riemannian manifold and
let
\begin{equation}
 K=K^\mu(X)\partial_\mu
\end{equation}
be a vector field on $\mathcal{M}$.  The vector field defines a local flow
through its integral curves,
\begin{equation}
 \frac{\dd X^\mu(\lambda)}{\dd\lambda}
 =
 K^\mu\bigl(X(\lambda)\bigr),
 \label{eq:app-flow}
\end{equation}
where $\lambda$ parametrizes motion along the orbit generated by $K$.
Equivalently, $K$ generates a one-parameter family of local maps
$\Phi_\lambda:\mathcal{M}\rightarrow\mathcal{M}$.  The infinitesimal change
of the metric along this flow is measured by its Lie derivative,
\begin{align}
 (\mathcal{L}_K g)_{\mu\nu}
 &=
 K^\rho\partial_\rho g_{\mu\nu}
 +
 g_{\rho\nu}\partial_\mu K^\rho
 +
 g_{\mu\rho}\partial_\nu K^\rho
 \nonumber\\
 &=
 \nabla_\mu K_\nu+\nabla_\nu K_\mu .
 \label{eq:app-Lie-derivative}
\end{align}

The distinction between Killing, homothetic and conformal Killing vector
fields is entirely determined by how the metric changes along this flow.

\subsection*{Killing, homothetic and conformal Killing flows.}

A Killing vector satisfies
\begin{equation}
 \mathcal{L}_K g_{\mu\nu}=0.
 \label{eq:app-Killing}
\end{equation}
Its flow therefore generates an isometry of the target space: the metric is
preserved exactly.

A conformal Killing vector satisfies
\begin{equation}
 \mathcal{L}_K g_{\mu\nu}
 =
 2\omega_K(X)g_{\mu\nu},
 \label{eq:app-CKV}
\end{equation}
where
\begin{equation}
 \omega_K(X)
 =
 \frac{1}{D}\nabla_\mu K^\mu .
 \label{eq:app-omega}
\end{equation}
Thus the metric is preserved only up to a local rescaling.  Although lengths
need not be preserved, the conformal structure is unchanged.  In
particular, null directions are preserved: if
\begin{equation}
 g_{\mu\nu}v^\mu v^\nu=0,
\end{equation}
then the same vector remains null after a conformal rescaling of the metric.

A homothetic vector is the special case for which the conformal factor is
constant,
\begin{equation}
 \mathcal{L}_\rho g_{\mu\nu}
 =
 2c\,g_{\mu\nu},
 \qquad c=\text{constant}.
 \label{eq:app-homothety-general}
\end{equation}
For a proper homothety $c\neq0$.  After rescaling the normalization of
$\rho$, we set $c=1$, so that
\begin{equation}
 \mathcal{L}_\rho g_{\mu\nu}
 =
 2g_{\mu\nu},
 \qquad
 \nabla_\mu\rho_\nu+\nabla_\nu\rho_\mu
 =
 2g_{\mu\nu}.
 \label{eq:app-normalized-homothety}
\end{equation}
Taking the trace gives
\begin{equation}
 \nabla_\mu\rho^\mu=D.
 \label{eq:app-homothety-divergence}
\end{equation}

The three possibilities can therefore be summarized as follows
\begin{equation}
 \begin{array}{|c|c|c|}
 \hline
 \text{Vector field}
 &
 \mathcal{L}_K g_{\mu\nu}
 &
 \text{Geometric meaning}
 \\ \hline
 \text{Killing}
 &
 0
 &
 \text{Isometry}
 \\[1mm]
 \text{Homothetic}
 &
 2c\,g_{\mu\nu}
 &
 \text{Uniform rescaling}
 \\[1mm]
 \text{Generic CKV}
 &
 2\omega_K(X)g_{\mu\nu}
 &
 \text{Position-dependent rescaling}\\
 \hline
 \end{array}
 \label{eq:app-vector-summary}
\end{equation}
Thus Killing and homothetic vectors are special cases of the conformal
Killing equation, distinguished by the conformal factor.

\subsection*{Relevance for Carroll-Weyl gauging.}

The distinction above becomes essential once the target-space flow itself
is gauged.  In the minimal Carroll-Weyl construction the embedding
coordinates transform along a target-space vector field $\rho$,
\begin{equation}
 \delta_\chi X^\mu
 =
 \chi\,\rho^\mu(X),
 \qquad
 \delta_\chi V^a=-\chi V^a .
 \label{eq:app-CW-minimal}
\end{equation}
Since
\begin{equation}
 \delta_\chi(V^aV^b)
 =
 -2\chi V^aV^b,
\end{equation}
the target-space kinetic term must acquire the opposite weight.

This happens naturally when $\rho$ is a normalized homothetic vector,
\begin{equation}
 \mathcal{L}_\rho g_{\mu\nu}=2g_{\mu\nu}.
\end{equation}
The metric then acquires precisely the scaling required to compensate the
Carroll-Weyl weight of $V^aV^b$.  A proper homothety is therefore the
natural geometric input for the minimal theory.

An ordinary Killing vector,
\begin{equation}
 \mathcal{L}_K g_{\mu\nu}=0,
\end{equation}
does not provide this compensating rescaling.  Killing vectors may of course
remain global target-space symmetries of the background, but within the
gauging considered here they are not the appropriate generators of the
Carroll-Weyl scaling.

The more general possibility is a conformal Killing vector,
\begin{equation}
 \mathcal{L}_K G_{\mu\nu}
 =
 2\omega_K(X)G_{\mu\nu}.
 \label{eq:app-generic-CKV}
\end{equation}
The corresponding nearest-to-minimal construction remains gauge invariant
if the worldsheet kernel transforms as
\begin{equation}
 \delta_\chi E^a
 =
 -\chi\omega_K(X)E^a .
 \label{eq:app-nearest-transform}
\end{equation}
For $\omega_K=1$ this reduces to the minimal theory.  For a generic CKV the
Carroll-Weyl weight depends explicitly on the target-space embedding.

Thus the relevant hierarchy for curved Carroll-Weyl strings is
\begin{equation*}
 \begin{array}{ccl}
 \text{Homothety}
 &:&
 \text{Minimal Carroll-Weyl theory},
 \\[1mm]
 \text{Generic CKV}
 &:&
 \text{Nearest-to-minimal theory},
 \\[1mm]
 \text{Killing vector}
 &:&
 \text{No Carroll-Weyl metric scaling}.
 \end{array}
 \label{eq:app-CW-hierarchy}
\end{equation*}

\subsection*{Representative backgrounds: Minkowski, AdS, dS.}

The distinction between Killing, homothetic and conformal Killing flows is
particularly transparent in a few standard backgrounds that are relevant for
the Carroll-Weyl construction.

\paragraph{Minkowski spacetime.}
For $D$-dimensional Minkowski spacetime,
\begin{equation}
 ds^2=\eta_{\mu\nu}\dd X^\mu\dd X^\nu,
\end{equation}
the Euler vector
\begin{equation}
 \rho=X^\mu\partial_\mu
\end{equation}
generates the dilation flow
\begin{equation}
 X^\mu\longrightarrow e^\lambda X^\mu
\end{equation}
and satisfies
\begin{equation}
 \mathcal{L}_{X\cdot\partial}\eta_{\mu\nu}
 =
 2\eta_{\mu\nu}.
 \label{eq:app-flat-homothety}
\end{equation}
It is therefore a proper normalized homothety and provides the natural
target-space flow for the minimal Carroll-Weyl gauging.  Minkowski space
also admits the usual Killing vectors generating translations and Lorentz
transformations, as well as generic conformal Killing vectors, but the
dilation vector is distinguished by its constant nonzero conformal factor.

\paragraph{Anti-de Sitter spacetime.}
Consider AdS$_D$ in the Poincar\'e patch,
\begin{equation}
 ds_G^2
 =
 \frac{L^2}{z^2}
 \left(
   \dd z^2+\eta_{ij}\dd x^i\dd x^j
 \right).
 \label{eq:app-AdS-metric}
\end{equation}
The radial vector
\begin{equation}
 K=\partial_z
\end{equation}
obeys
\begin{equation}
 \mathcal{L}_{\partial_z}G_{\mu\nu}
 =
 -\frac{2}{z}G_{\mu\nu},
 \label{eq:app-AdS-CKV}
\end{equation}
so that
\begin{equation}
 \omega_K=-\frac{1}{z}.
\end{equation}
Hence $\partial_z$ is a generic conformal Killing vector, rather than a
homothety, of the physical AdS metric.  It should be distinguished from
\begin{equation}
 z\partial_z+x^i\partial_i,
\end{equation}
which generates an exact AdS isometry.  Although the latter resembles a
coordinate dilation, its Lie derivative of the physical metric vanishes and
it therefore does not provide the nonzero metric scaling required for the
Carroll-Weyl gauging considered here.

\paragraph{de Sitter spacetime.}
In the cosmological patch of de Sitter space,
\begin{equation}
 ds_G^2
 =
 \frac{1}{H^2\eta^2}
 \left(
   -\dd\eta^2+\delta_{ij}\dd x^i\dd x^j
 \right),
\end{equation}
the natural time-directed vector
\begin{equation}
 K=\partial_\eta
\end{equation}
satisfies
\begin{equation}
 \mathcal{L}_{\partial_\eta}G_{\mu\nu}
 =
 -\frac{2}{\eta}G_{\mu\nu}.
\end{equation}
Thus $\partial_\eta$ is again a generic conformal Killing vector with a
position-dependent conformal factor, rather than a homothety of the physical
dS metric.

A useful common feature of AdS and dS is that neither physical metric admits
a proper homothety.  Suppose
\begin{equation}
 \mathcal{L}_\rho g_{\mu\nu}
 =
 2c\,g_{\mu\nu},
 \qquad
 c=\text{constant}.
\end{equation}
The scalar curvature then transforms as
\begin{equation}
 \mathcal{L}_\rho R=-2cR.
\end{equation}
Since both AdS and dS have nonzero constant scalar curvature, one also has
$\mathcal{L}_\rho R=0$, and therefore $c=0$.  Hence any homothetic vector of
the physical AdS or dS metric is necessarily Killing.  This is why these
backgrounds naturally lead to the conformal-Killing-vector, rather than the
proper-homothety, version of the Carroll-Weyl construction.

\subsection*{From a CKV to a homothetic frame.}

The generic CKV formulation is sufficient for the nearest-to-minimal
Carroll-Weyl construction.  It is nevertheless often convenient to pass to
a conformally related frame in which the same target-space flow becomes
homothetic.  This brings the Carroll-Weyl transformation to the simpler
fixed-weight form of the minimal theory.

Suppose that $K$ is a conformal Killing vector of the physical metric
$G_{\mu\nu}$,
\begin{equation}
 \mathcal{L}_K G_{\mu\nu}
 =
 2\omega_K(X)G_{\mu\nu}.
 \label{eq:app-physical-CKV}
\end{equation}
Consider a conformally related metric
\begin{equation}
 g_{\mu\nu}
 =
 e^{-2\Omega(X)}G_{\mu\nu}.
 \label{eq:app-conformal-frame}
\end{equation}
Its Lie derivative along the same vector field is
\begin{equation}
 \mathcal{L}_K g_{\mu\nu}
 =
 2\left(
 \omega_K-K^\rho\partial_\rho\Omega
 \right)g_{\mu\nu}.
 \label{eq:app-frame-lie}
\end{equation}
Therefore, if the conformal factor is chosen to satisfy
\begin{equation}
 K^\rho\partial_\rho\Omega
 =
 \omega_K-1,
 \label{eq:app-frame-equation}
\end{equation}
then
\begin{equation}
 \mathcal{L}_K g_{\mu\nu}
 =
 2g_{\mu\nu}.
 \label{eq:app-K-homothetic}
\end{equation}
Thus the same vector field $K$, which is a generic CKV of the physical
metric $G_{\mu\nu}$, becomes a normalized homothetic vector of the
conformally related metric $g_{\mu\nu}$.

Equation~\eqref{eq:app-frame-equation} is a first-order differential equation
along the integral curves of $K$ and can be solved locally on a regular
patch where $K\neq0$.  Importantly, this conformal change of frame does not
alter the target-space flow being gauged; it only chooses a representative
of the conformal class in which that flow acts with a constant scaling
weight.

The homothetic frame is therefore particularly useful for the
Carroll-Weyl description: the position-dependent weight
$\omega_K(X)$ of the nearest-to-minimal formulation is replaced by the
constant unit weight of the minimal theory.  This is precisely the strategy
used for the AdS Poincar\'e patch in the main text, where the radial CKV of
the physical AdS metric is mapped to a homothety of a suitable conformal
representative.

\section{The radial Carroll-Weyl constraint}
\label{app:radial-constraint-checks}

In \cref{subsec:ads-phase-space} the Carroll-Weyl constraint associated
with the radial flow
\begin{equation}
 K=\partial_z
\end{equation}
was identified as
\begin{equation}
 \cS=P_z\approx0.
 \label{eq:app-Pz-constraint}
\end{equation}
Here we verify this statement directly from the canonical phase-space
description.

The nonvanishing equal-time canonical brackets are
\begin{equation}
 \{z(\sigma),P_z(\sigma')\}
 =
 2\pi\delta(\sigma-\sigma'),
 \qquad
 \{x^i(\sigma),P_j(\sigma')\}
 =
 2\pi\delta^i{}_j\delta(\sigma-\sigma').
 \label{eq:app-canonical-brackets}
\end{equation}
Consider the smeared generator
\begin{equation}
 S[\chi]
 =
 \frac{1}{2\pi}
 \int\dd\sigma'\,
 \chi(\sigma')P_z(\sigma').
 \label{eq:app-smeared-Pz}
\end{equation}
Its action on the radial coordinate is
\begin{align}
 \delta_\chi z(\sigma)
 &=
 \{z(\sigma),S[\chi]\}
 \nonumber\\
 &=
 \frac{1}{2\pi}
 \int\dd\sigma'\,
 \chi(\sigma')
 \{z(\sigma),P_z(\sigma')\}
 \nonumber\\
 &=
 \chi(\sigma).
 \label{eq:app-Pz-z-shift}
\end{align}
Thus
\begin{equation}
 \delta_\chi z=\chi,
 \label{eq:app-z-transform}
\end{equation}
which is precisely the local radial shift generated by the vector field
$K=\partial_z$.

The transverse coordinates are invariant:
\begin{align}
 \delta_\chi x^i(\sigma)
 &=
 \{x^i(\sigma),S[\chi]\}
 \nonumber\\
 &=
 \frac{1}{2\pi}
 \int\dd\sigma'\,
 \chi(\sigma')
 \{x^i(\sigma),P_z(\sigma')\}=0.
 \label{eq:app-x-transform}
\end{align}
Hence
\begin{equation}
 \delta_\chi x^i=0.
\end{equation}
The same generator also acts trivially on the canonical momenta,
\begin{equation}
 \delta_\chi P_z=0,
 \qquad
 \delta_\chi P_i=0.
 \label{eq:app-momenta-transform}
\end{equation}

Therefore $P_z$ generates a pure local translation along the radial
coordinate, leaving the transverse phase-space variables unchanged.  This
is exactly the canonical action of the target-space vector field
$K=\partial_z$.

More generally, if a target-space flow is generated by a vector field
$K=K^A(X)\partial_A$, its canonical moment map is
\begin{equation}
 \cS_K=K^A(X)P_A.
 \label{eq:app-general-SK}
\end{equation}
Indeed, for the smeared generator
\begin{equation}
 S_K[\chi]
 =
 \frac{1}{2\pi}
 \int\dd\sigma\,
 \chi(\sigma)K^A(X)P_A,
\end{equation}
we find
\begin{equation}
 \delta_\chi X^A
 =
 \{X^A,S_K[\chi]\}
 =
 \chi K^A(X),
 \label{eq:app-general-K-action}
\end{equation}
so that $\cS_K$ generates the local flow along $K$.

For the AdS Poincar\'e realization considered in the main text,
\begin{equation}
 K^A\partial_A=\partial_z,
 \qquad
 K^z=1,
 \qquad
 K^i=0,
\end{equation}
and therefore
\begin{equation}
 \cS_K
 =
 K^AP_A
 =
 P_z.
\end{equation}
The Carroll-Weyl constraint associated with the radial flow is consequently
\begin{equation}
 \cS=P_z\approx0.
\end{equation}
This also makes clear why the AdS radial generator differs from the flat
Cartesian Carroll-Weyl generator
$S_{\rm flat}=X^AP_A$; the former generates a radial translation, whereas
the latter generates a target-space dilation.

\section{Details of the radial AdS residual algebra}
\label{app:ads-residual-algebra}

In this appendix we derive the residual constraint algebra of the
Carroll-Weyl gauged null string in the AdS Poincar\'e patch directly from
the canonical phase-space brackets and summarize its algebraic structure.
The derivation also fixes the smearing and Fourier-mode conventions used in
the main text.

We use the collective notation
\begin{equation}
 X^A=(z,x^i),
 \qquad
 P_A=(P_z,P_i),
\end{equation}
for which the three constraint densities are
\begin{equation}
 \cM=\frac12 P_AP^A,
 \qquad
 \cH=P_AX'^A,
 \qquad
 \cS=P_z.
 \label{eq:app-ads-constraints}
\end{equation}

The equal-time canonical brackets are
\begin{equation}
 \{X^A(\sigma),P_B(\sigma')\}
 =
 2\pi\delta^A{}_B\,\delta(\sigma-\sigma'),
 \label{eq:app-canonical-brackets}
\end{equation}
with
\begin{equation}
 \{X^A(\sigma),X^B(\sigma')\}=0,
 \qquad
 \{P_A(\sigma),P_B(\sigma')\}=0.
\end{equation}

\subsection*{Smeared constraint algebra.}

Introduce the normalized smeared generators
\begin{align}
 H[f]
 &=
 \frac{1}{2\pi}\int\dd\sigma\,
 f(\sigma)\cH(\sigma)
 =
 \frac{1}{2\pi}\int\dd\sigma\,fP_AX'^A,
 \nonumber\\
 M[g]
 &=
 \frac{1}{2\pi}\int\dd\sigma\,
 g(\sigma)\cM(\sigma)
 =
 \frac{1}{4\pi}\int\dd\sigma\,gP_AP^A,
 \nonumber\\
 S[h]
 &=
 \frac{1}{2\pi}\int\dd\sigma\,
 h(\sigma)\cS(\sigma)
 =
 \frac{1}{2\pi}\int\dd\sigma\,hP_z.
 \label{eq:app-smeared-generators}
\end{align}
We assume a closed spatial circle, so total $\sigma$-derivatives can be
discarded.

The generator $H[f]$ acts as the spatial reparametrization generator.  From
\eqref{eq:app-canonical-brackets},
\begin{align}
 \{X^A(\sigma),H[f]\}
 &=
 \frac{1}{2\pi}
 \int\dd\sigma'\,
 f(\sigma')
 \{X^A(\sigma),
 P_B(\sigma')X'^B(\sigma')\}
 \nonumber\\
 &=
 f(\sigma)X'^A(\sigma),
\end{align}
and therefore
\begin{equation}
 \{X^A,H[f]\}=fX'^A.
 \label{eq:app-X-under-H}
\end{equation}
Similarly,
\begin{align}
 \{P_A(\sigma),H[f]\}
 &=
 \frac{1}{2\pi}
 \int\dd\sigma'\,
 f(\sigma')P_B(\sigma')
 \{P_A(\sigma),X'^B(\sigma')\}
 \nonumber\\
 &=
 -\int\dd\sigma'\,
 f(\sigma')P_A(\sigma')
 \partial_{\sigma'}
 \delta(\sigma-\sigma')
 \nonumber\\
 &=
 \partial_\sigma
 \bigl(f(\sigma)P_A(\sigma)\bigr),
\end{align}
so that
\begin{equation}
 \{P_A,H[f]\}=(fP_A)'.
 \label{eq:app-P-under-H}
\end{equation}

Equations~\eqref{eq:app-X-under-H} and
\eqref{eq:app-P-under-H} immediately determine the transformation
properties of the three constraint densities.  For $\cM$,
\begin{align}
 \{\cM,H[f]\}
 &=
 P^A\{P_A,H[f]\}
 \nonumber\\
 &=
 P^A(fP_A)'
 \nonumber\\
 &=
 f\cM'+2f'\cM,
\end{align}
and hence
\begin{equation}
 \{\cM,H[f]\}=f\cM'+2f'\cM.
 \label{eq:app-M-transform}
\end{equation}
Thus $\cM$ transforms as a weight-two density.

Since $\cS=P_z$,
\begin{equation}
 \{\cS,H[f]\}
 =
 f\cS'+f'\cS,
 \label{eq:app-S-transform}
\end{equation}
so $\cS$ has weight one.  Finally,
\begin{align}
 \{\cH,H[f]\}
 &=
 \{P_AX'^A,H[f]\}
 \nonumber\\
 &=
 (fP_A)'X'^A
 +
 P_A(fX'^A)'
 \nonumber\\
 &=
 f\cH'+2f'\cH,
\end{align}
and $\cH$ is also a weight-two density.

The smeared algebra now follows by integrating these transformation laws.
For example,
\begin{align}
 \{M[g],H[f]\}
 &=
 \frac{1}{2\pi}
 \int\dd\sigma\,
 g(f\cM'+2f'\cM)
 \nonumber\\
 &=
 M[gf'-fg'] ,
\end{align}
and therefore
\begin{equation}
 \{H[f],M[g]\}
 =
 M[fg'-f'g].
\end{equation}
The same argument gives
\begin{equation}
 \{H[f],H[g]\}
 =
 H[fg'-f'g].
\end{equation}

For the radial constraint,
\begin{align}
 \{S[h],H[f]\}
 &=
 \frac{1}{2\pi}
 \int\dd\sigma\,
 h(f\cS'+f'\cS)
 \nonumber\\
 &=
 -S[fh'],
\end{align}
and hence
\begin{equation}
 \{H[f],S[h]\}
 =
 S[fh'].
\end{equation}

The remaining brackets vanish strongly.  Since $\cM$ depends only on
mutually commuting momenta,
\begin{equation}
 \{M[f],M[g]\}=0.
\end{equation}
Likewise,
\begin{equation}
 \{S[h],S[k]\}=0.
\end{equation}
Most importantly,
\begin{align}
 \{\cS(\sigma),\cM(\sigma')\}
 &=
 \left\{
 P_z(\sigma),
 \frac12
 \bigl(P_z^2+P_iP^i\bigr)(\sigma')
 \right\}
 \nonumber\\
 &=0,
\end{align}
so that
\begin{equation}
 \{S[h],M[g]\}=0.
\end{equation}

Collecting the results,
\begin{align}
 \{H[f],H[g]\}
 &=
 H[fg'-f'g],
 &
 \{H[f],M[g]\}
 &=
 M[fg'-f'g],
 \nonumber\\
 \{H[f],S[h]\}
 &=
 S[fh'],
 &
 \{S[h],M[g]\}
 &=
 0,
 \nonumber\\
 \{M[f],M[g]\}
 &=
 0,
 &
 \{S[h],S[k]\}
 &=
 0.
 \label{eq:app-full-smeared-algebra}
\end{align}

The vanishing of the $SM$ bracket is a characteristic feature of the AdS
radial realization.  Here $\cS=P_z$ generates translations along the
distinguished radial direction, while $\cM$ contains no explicit dependence
on $z$.  The two constraints therefore commute strongly.  This differs from
the Cartesian Carroll-Weyl realization, in which the third generator is a
target-space dilation and acts nontrivially on the null Hamiltonian.

\subsection*{Fourier modes and the residual algebra.}

To pass to Fourier modes, take
\begin{equation}
 f_m(\sigma)=e^{im\sigma},
 \qquad
 f'_m=imf_m,
\end{equation}
and define
\begin{equation}
 H_m=H[f_m],
 \qquad
 M_m=M[f_m],
 \qquad
 S_m=S[f_m].
 \label{eq:app-mode-definitions}
\end{equation}
Since
\begin{equation}
 f_mf'_n-f'_mf_n
 =
 i(n-m)e^{i(m+n)\sigma},
\end{equation}
the smeared brackets give
\begin{equation}
 \{H_m,H_n\}
 =
 i(n-m)H_{m+n},
\end{equation}
and
\begin{equation}
 \{H_m,M_n\}
 =
 i(n-m)M_{m+n}.
\end{equation}
For the weight-one current,
\begin{equation}
 \{H_m,S_n\}
 =
 in\,S_{m+n}.
\end{equation}
The remaining brackets vanish:
\begin{equation}
 \{S_m,M_n\}=0,
 \qquad
 \{M_m,M_n\}=0,
 \qquad
 \{S_m,S_n\}=0.
\end{equation}

Using the convention
\begin{equation}
 [\,\cdot,\cdot\,]
 =
 i\{\,\cdot,\cdot\,\},
\end{equation}
one obtains
\begin{align}
 [H_m,H_n]
 &=
 (m-n)H_{m+n},
 &
 [H_m,M_n]
 &=
 (m-n)M_{m+n},
 \nonumber\\
 [H_m,S_n]
 &=
 -nS_{m+n},
 &
 [S_m,M_n]
 &=
 0,
 \nonumber\\
 [M_m,M_n]
 &=
 0,
 &
 [S_m,S_n]
 &=
 0.
 \label{eq:app-HMS-algebra}
\end{align}

In the Laurent-current convention used in the main text, the
spatial-reparametrization generator is denoted by $L_m$.  With the adopted
orientation and normalization of the Laurent modes, the residual algebra is
therefore
\begin{align}
 [L_m,L_n]
 &=
 (m-n)L_{m+n},
 &
 [L_m,M_n]
 &=
 (m-n)M_{m+n},
 \nonumber\\
 [L_m,S_n]
 &=
 -nS_{m+n},
 &
 [S_m,M_n]
 &=
 0,
 \nonumber\\
 [M_m,M_n]
 &=
 0,
 &
 [S_m,S_n]
 &=
 0.
 \label{eq:app-ads-residual-algebra}
\end{align}
This is the centerless residual algebra used throughout the quantum
analysis.

\subsection*{Algebraic interpretation.}

The structure of \eqref{eq:app-ads-residual-algebra} is straightforward.
The generators $L_n$ form the Witt algebra, while the $M_n$ and $S_n$
together form an abelian ideal on which the Witt generators act
nontrivially.  The two sectors transform differently: $M_n$ behaves as a
weight-two current, whereas $S_n$ behaves as a weight-one current.

The generators $L_n$ form the Witt algebra.  The generators $M_n$ and
$S_n$ form an abelian ideal on which the Witt generators act
nontrivially.  In particular,
\begin{equation}
 [L_m,M_n]=(m-n)M_{m+n},
\end{equation}
so $M_n$ transforms in the same way as a weight-two current, while
\begin{equation}
 [L_m,S_n]=-nS_{m+n},
\end{equation}
shows that $S_n$ transforms as a weight-one current.

In the standard conformal-mode convention,
\begin{equation}
 [L_m,\Phi_n]
 =
 \bigl((h-1)m-n\bigr)\Phi_{m+n},
\end{equation}
and therefore
\begin{equation}
 h_M=2,
 \qquad
 h_S=1.
\end{equation}
Thus the $(L_n,M_n)$ sector is the centerless $\mathfrak{bms}_3$ algebra,
while the additional $S_n$ generators form an abelian weight-one current
sector commuting with $M_n$.

It is therefore convenient to regard the full algebra as
\begin{equation}
 \mathfrak a_{\rm AdS}
 =
 \mathrm{Witt}
 \ltimes
 \left(
 \mathcal F_{-1}\oplus\mathcal F_0
 \right),
 \label{eq:app-density-decomposition}
\end{equation}
where $M_n\in\mathcal F_{-1}$ and $S_n\in\mathcal F_0$.  The notation
$\mathcal F_\lambda$ records the corresponding tensor-density module under
the Witt action.

The algebra is also naturally $\mathbb Z$-graded:
\begin{equation}
 \mathfrak a_{\rm AdS}
 =
 \bigoplus_{n\in\mathbb Z}
 \operatorname{span}\{L_n,M_n,S_n\}.
\end{equation}
The grading is generated by $L_0$,
\begin{equation}
 [L_0,L_n]=-nL_n,
 \qquad
 [L_0,M_n]=-nM_n,
 \qquad
 [L_0,S_n]=-nS_n.
 \label{eq:app-L0-grading}
\end{equation}
In particular,
\begin{equation}
 [L_m,S_0]
 =
 [M_m,S_0]
 =
 [S_m,S_0]
 =
 0,
\end{equation}
so $S_0$ is central in the centerless algebra.

The decomposition
\eqref{eq:app-density-decomposition} and the grading
\eqref{eq:app-L0-grading} will be useful in the Chevalley--Eilenberg
analysis of the possible central extensions.

\section{Lie algebra cohomology of the radial AdS residual algebra}
\label{app:adscohomology}

This appendix determines the central extensions of the classical residual
algebra of the Carroll-Weyl gauged null string in the AdS Poincar\'e patch.
The discussion is purely algebraic.  In particular, the possible central
extensions are determined before choosing a quantum realization or computing
operator-product expansions.

The classical modes obey
\begin{align}
 [L_m,L_n]&=(m-n)L_{m+n},&
 [L_m,M_n]&=(m-n)M_{m+n}, \nonumber\\
 [L_m,S_n]&=-nS_{m+n},&
 [M_m,M_n]&=[M_m,S_n]=[S_m,S_n]=0 .
 \label{eq:adsCWclassical}
\end{align}
In the tensor-density convention \cite{OvsienkoRoger1996,Ovsienko:1998TensorDensity},
\begin{equation}
 \mathfrak a_{\rm AdS}
 =\mathrm{Witt}\ltimes\left(\mathcal F_{-1}\oplus\mathcal F_{0}\right),
 \label{eq:adsCWsemidirect}
\end{equation}
where $M_n\in\mathcal F_{-1}$ and $S_n\in\mathcal F_0$.
The semidirect-product structure can equivalently be expressed through
the split short exact sequence
\begin{equation}
0
\longrightarrow
\mathcal F_{-1}\oplus\mathcal F_0
\xrightarrow{\;\iota\;}
\mathfrak a_{\rm AdS}
\xrightarrow{\;\pi\;}
\mathrm{Witt}
\longrightarrow
0 .
\label{eq:ads-short-exact}
\end{equation}
Here $\mathcal F_{-1}\oplus\mathcal F_0$ is the abelian ideal generated
by $\{M_n,S_n\}$, $\iota$ denotes its natural inclusion, and $\pi$
projects onto the Witt sector,
\begin{equation}
\pi(L_n)=L_n,
\qquad
\pi(M_n)=\pi(S_n)=0.
\end{equation}
Thus
\begin{equation}
\ker\pi
=
\operatorname{im}\iota
=
\mathcal F_{-1}\oplus\mathcal F_0.
\end{equation}
The sequence is split by the homomorphism
$s:\mathrm{Witt}\rightarrow\mathfrak a_{\rm AdS}$ defined by
$s(L_n)=L_n$, for which $\pi\circ s=\mathrm{id}_{\mathrm{Witt}}$.
This splitting is precisely the algebraic content of
$\mathfrak a_{\rm AdS}
=\mathrm{Witt}\ltimes(\mathcal F_{-1}\oplus\mathcal F_0)$.

A central extension is specified by an antisymmetric bilinear map
$\gamma:\mathfrak a_{\rm AdS}\wedge\mathfrak a_{\rm AdS}\to\mathbb C$
obeying the Chevalley-Eilenberg cocycle condition
\begin{equation}
 \gamma([X,Y],Z)+\gamma([Y,Z],X)+\gamma([Z,X],Y)=0.
 \label{eq:CEcondition}
\end{equation}
Using the $L_0$ grading, we choose a representative of each cohomology
class of degree zero.  The most general such $2$-cochain is
\begin{align}
 \gamma(L_m,L_n)&=A(m)\delta_{m+n,0},&
 \gamma(L_m,M_n)&=B(m)\delta_{m+n,0},\nonumber\\
 \gamma(L_m,S_n)&=C(m)\delta_{m+n,0},&
 \gamma(M_m,M_n)&=D(m)\delta_{m+n,0},\nonumber\\
 \gamma(M_m,S_n)&=E(m)\delta_{m+n,0},&
 \gamma(S_m,S_n)&=F(m)\delta_{m+n,0}.
 \label{eq:general2cochain}
\end{align}
Antisymmetry requires
\begin{equation}
 A(-m)=-A(m),\qquad
 D(-m)=-D(m),\qquad
 F(-m)=-F(m).
 \label{eq:antisymmetry}
\end{equation}

The $(L_m,L_n,L_p)$ cocycle condition gives
\begin{equation}
 (m-n)A(m+n)+(n-p)A(n+p)+(p-m)A(p+m)=0,
 \qquad m+n+p=0.
 \label{eq:AAAcondition}
\end{equation}
Its solution is $A(m)=a_3m^3+a_1m$.  The term proportional to $m$ is a
coboundary, so a representative of the nontrivial class can be chosen as
\begin{equation}
 \gamma_{LL}(L_m,L_n)
 =\frac{1}{12}m(m^2-1)\delta_{m+n,0}.
 \label{eq:gammaLLAdS}
\end{equation}

The $(L_m,L_n,M_p)$ cocycle condition yields
\begin{equation}
 (m-n)B(m+n)-(m+2n)B(m)+(2m+n)B(n)=0,
 \qquad m+n+p=0.
 \label{eq:BBBcondition}
\end{equation}
Hence $B(m)=b_3m^3+b_1m$.  The linear term is a coboundary, and the
nontrivial mixed BMS representative can be written as
\begin{equation}
 \gamma_{LM}(L_m,M_n)
 =\frac{1}{12}m(m^2-1)\delta_{m+n,0}.
 \label{eq:gammaLMAdS}
\end{equation}

Similarly, the $(L_m,L_n,S_p)$ condition is
\begin{equation}
 (m-n)C(m+n)-(m+n)\bigl(C(m)-C(n)\bigr)=0,
 \qquad m+n+p=0.
 \label{eq:CCCcondition}
\end{equation}
It gives $C(m)=c_2m^2+c_1m$, where the linear piece is again a coboundary.
A convenient representative is
\begin{equation}
 \gamma_{LS}(L_m,S_n)
 =\frac12m(m+1)\delta_{m+n,0}.
 \label{eq:gammaLSAdS}
\end{equation}

For the $SS$ sector, the $(L_k,S_m,S_n)$ identity becomes
\begin{equation}
 -mF(k+m)+(k+m)F(m)=0,
 \label{eq:FFFcondition}
\end{equation}
and therefore
\begin{equation}
 \gamma_{SS}(S_m,S_n)
 =m\delta_{m+n,0}.
 \label{eq:gammaSSAdS}
\end{equation}

There are no independent $MM$ or $MS$ central extensions.  Indeed, the
$(L_k,M_m,M_n)$ and $(L_k,M_m,S_n)$ cocycle conditions imply, respectively,
\begin{align}
 (k-m)D(k+m)+(k-n)D(m)&=0,
 \qquad n=-k-m, \label{eq:MMcondition}\\
 (k-m)E(k+m)+(k+m)E(m)&=0. \label{eq:MScondition}
\end{align}
These equations give
\begin{equation}
 D(m)=0,\qquad E(m)=0.
 \label{eq:MMMSvanish}
\end{equation}

Consequently,
\begin{equation}
 H^2(\mathfrak a_{\rm AdS};\mathbb C)
 \simeq\mathbb C^4,
 \label{eq:H2AdS}
\end{equation}
with representatives
\begin{equation}
 \left\{
 \gamma_{LL},\,
 \gamma_{LM},\,
 \gamma_{LS},\,
 \gamma_{SS}
 \right\}.
 \label{eq:AdScocyclebasis}
\end{equation}
The most general centrally extended algebra is therefore
\begin{align}
 [L_m,L_n]&=(m-n)L_{m+n}
 +\frac{c_{LL}}{12}m(m^2-1)\delta_{m+n,0}, \nonumber\\
 [L_m,M_n]&=(m-n)M_{m+n}
 +\frac{c_{LM}}{12}m(m^2-1)\delta_{m+n,0}, \nonumber\\
 [L_m,S_n]&=-nS_{m+n}
 +\frac{c_{LS}}{2}m(m+1)\delta_{m+n,0}, \nonumber\\
 [S_m,S_n]&=c_{SS}\,m\delta_{m+n,0}.
 \label{eq:AdScentrallyextended}
\end{align}

Equation~\eqref{eq:H2AdS} is a statement about the abstract residual gauge
algebra.  A choice of matter fields, ghosts, measure, regulator, and normal
ordering realizes a particular point
\begin{equation}
 (c_{LL},c_{LM},c_{LS},c_{SS})\in\mathbb C^4.
 \label{eq:AdSlevels}
\end{equation}
Thus, if a particular free-field realization yields
\begin{equation}
 (c_{LL},c_{LM},c_{LS},c_{SS})=(2D-54,0,0,0),
 \label{eq:canonicalAdSlevels}
\end{equation}
the three vanishing entries are properties of that realization.  They do not
mean that the $LM$, $LS$, and $SS$ cocycles are absent cohomologically.


\subsection*{Chevalley-Eilenberg cochain complex and its degree.}
Let $\mathfrak g$ be a Lie algebra and let $V$ be a $\mathfrak g$-module. The
Chevalley-Eilenberg cochain complex of $\mathfrak g$ with coefficients in
$V$ is the sequence
\begin{equation}
0
\longrightarrow
C^0(\mathfrak g;V)
\xrightarrow{\,d_{\rm CE}\,}
C^1(\mathfrak g;V)
\xrightarrow{\,d_{\rm CE}\,}
C^2(\mathfrak g;V)
\xrightarrow{\,d_{\rm CE}\,}
C^3(\mathfrak g;V)
\longrightarrow \cdots ,
\end{equation}
where
\begin{equation}
C^p(\mathfrak g;V)
=
\operatorname{Hom}\!\left(
\Lambda^p\mathfrak g,V
\right)
\end{equation}
is the space of antisymmetric $p$-linear maps from $\mathfrak g$ to $V$.
An element
\begin{equation}
\small
\omega\in C^p(\mathfrak g;V)
\end{equation}
is called a $p$-cochain.

The Chevalley-Eilenberg differential
\begin{equation}
\small
d_{\rm CE}:
C^p(\mathfrak g;V)
\longrightarrow
C^{p+1}(\mathfrak g;V)
\end{equation}
is defined by
\begin{align}
\small
(d_{\rm CE}\omega)(X_0,\ldots,X_p)
={}&
\sum_{i=0}^{p}
(-1)^i
X_i\!\cdot\!
\omega(X_0,\ldots,\widehat X_i,\ldots,X_p)
\nonumber\\
&+
\sum_{0\leq i<j\leq p}
(-1)^{i+j}
\omega\!\left(
[X_i,X_j],
X_0,\ldots,
\widehat X_i,\ldots,
\widehat X_j,\ldots,
X_p
\right),
\end{align}
where $\widehat X_i$ denotes omission of the corresponding argument.
The Jacobi identity together with the compatibility of the
$\mathfrak g$-action on $V$ implies
\begin{equation}
d_{\rm CE}^{\,2}=0.
\end{equation}
It is this nilpotency property that makes the above sequence a cochain
complex.

In the present application we are interested in trivial coefficients,
$V=\mathbb C$. The Lie algebra acts trivially on $\mathbb C$, so that the
first term in the Chevalley-Eilenberg differential vanishes. Hence
\begin{equation}
C^p(\mathfrak g;\mathbb C)
=
\operatorname{Hom}
\!\left(\Lambda^p\mathfrak g,\mathbb C\right),
\end{equation}
and the differential is determined entirely by the Lie bracket,
\begin{equation}
(d_{\rm CE}\omega)(X_0,\ldots,X_p)
=
\sum_{i<j}
(-1)^{i+j}
\omega\!\left(
[X_i,X_j],
X_0,\ldots,
\widehat X_i,\ldots,
\widehat X_j,\ldots,
X_p
\right).
\end{equation}

A cochain $\omega\in C^p(\mathfrak g;\mathbb C)$ is called
\emph{closed}, or a $p$-cocycle, when
\begin{equation}
d_{\rm CE}\omega=0,
\end{equation}
and it is called \emph{exact}, or a $p$-coboundary, when there exists
$\eta\in C^{p-1}(\mathfrak g;\mathbb C)$ such that
\begin{equation}
\omega=d_{\rm CE}\eta.
\end{equation}
Since $d_{\rm CE}^{\,2}=0$, every coboundary is automatically a cocycle.
The Chevalley-Eilenberg cohomology is therefore the quotient
\begin{equation}
H^p(\mathfrak g;\mathbb C)
=
\frac{
\ker\!\left(
d_{\rm CE}:C^p\rightarrow C^{p+1}
\right)}
{
\operatorname{im}\!\left(
d_{\rm CE}:C^{p-1}\rightarrow C^p
\right)}
=
\frac{Z^p(\mathfrak g;\mathbb C)}
{B^p(\mathfrak g;\mathbb C)}.
\end{equation}

For the classification of central extensions, the relevant part of the
complex is
\begin{equation}
C^1(\mathfrak g;\mathbb C)
\xrightarrow{\,d_{\rm CE}\,}
C^2(\mathfrak g;\mathbb C)
\xrightarrow{\,d_{\rm CE}\,}
C^3(\mathfrak g;\mathbb C).
\end{equation}
A two-cochain is an antisymmetric bilinear map
\begin{equation}
\gamma:
\mathfrak g\wedge\mathfrak g
\longrightarrow
\mathbb C.
\end{equation}
For trivial coefficients, the condition that $\gamma$ be a two-cocycle
takes the form
\begin{equation}
\gamma([X,Y],Z)
+
\gamma([Y,Z],X)
+
\gamma([Z,X],Y)
=0.
\end{equation}
This is precisely the condition required for the centrally extended
bracket
\begin{equation}
[X,Y]_{\rm ext}
=
[X,Y]
+
\gamma(X,Y)\,\mathbf 1
\end{equation}
to satisfy the Jacobi identity.

Two cocycles which differ by a coboundary,
\begin{equation}
\gamma'
=
\gamma+d_{\rm CE}\eta,
\qquad
\eta\in C^1(\mathfrak g;\mathbb C),
\end{equation}
describe equivalent central extensions, since the difference may be
absorbed into a redefinition of the generators. Consequently,
inequivalent one-dimensional central extensions of $\mathfrak g$ are
classified by the second Chevalley-Eilenberg cohomology group
\begin{equation}
H^2(\mathfrak g;\mathbb C).
\end{equation}

Applying this construction to the AdS residual algebra
$\mathfrak a_{\rm AdS}$, we find
\begin{equation}
H^2(\mathfrak a_{\rm AdS};\mathbb C)
\simeq
\mathbb C^4,
\end{equation}
corresponding to four independent cohomology classes represented by the
$LL$, $LM$, $LS$, and $SS$ central-extension channels.

We now evaluate the cohomology at degree two of the Chevalley-Eilenberg cochain complex. For a Lie algebra
$\mathfrak g$ with trivial coefficients in $\mathbb C$, the degree-$p$
cochains are
\begin{equation}
 C^p(\mathfrak g;\mathbb C)
 =
 \operatorname{Hom}
 \bigl(\wedge^p\mathfrak g,\mathbb C\bigr),
\end{equation}
and the Chevalley-Eilenberg differential increases the degree by one,
\begin{equation}
 C^1(\mathfrak g;\mathbb C)
 \xrightarrow{\;\delta\;}
 C^2(\mathfrak g;\mathbb C)
 \xrightarrow{\;\delta\;}
 C^3(\mathfrak g;\mathbb C).
\end{equation}
A candidate central term is an antisymmetric bilinear map
\begin{equation}
 \gamma:
 \mathfrak a_{\rm AdS}\wedge
 \mathfrak a_{\rm AdS}
 \longrightarrow\mathbb C,
\end{equation}
and therefore belongs to
$C^2(\mathfrak a_{\rm AdS};\mathbb C)$.  The second cohomology is
defined as
\begin{equation}
 H^2(\mathfrak a_{\rm AdS};\mathbb C)
 =
 \frac{
 \ker\!\left(
 \delta:C^2\rightarrow C^3
 \right)}
 {
 \operatorname{im}\!\left(
 \delta:C^1\rightarrow C^2
 \right)}
 =
 \frac{Z^2(\mathfrak a_{\rm AdS};\mathbb C)}
      {B^2(\mathfrak a_{\rm AdS};\mathbb C)}.
\end{equation}
Thus the superscript $2$ reflects the fact that central extensions are
described by degree-two cochains, namely antisymmetric bilinear maps of
two Lie-algebra generators.

The second Chevalley-Eilenberg cohomology group of the residual algebra
with trivial coefficients is defined by
\begin{equation}
 H^2(\mathfrak a_{\rm AdS};\mathbb C)
 =
 \frac{Z^2(\mathfrak a_{\rm AdS};\mathbb C)}
      {B^2(\mathfrak a_{\rm AdS};\mathbb C)}
 =
 \frac{\{\text{2-cocycles}\}}
      {\{\text{2-coboundaries}\}} .
 \label{eq:H2-definition}
\end{equation}
Here a two-cochain is an antisymmetric bilinear map
\begin{equation}
 \gamma:
 \mathfrak a_{\rm AdS}\wedge
 \mathfrak a_{\rm AdS}
 \longrightarrow \mathbb C .
\end{equation}
It is a two-cocycle if it satisfies the
Chevalley-Eilenberg cocycle condition
\begin{equation}
 \gamma([X,Y],Z)
 +\gamma([Y,Z],X)
 +\gamma([Z,X],Y)
 =0 ,
 \label{eq:two-cocycle-condition}
\end{equation}
for all \(X,Y,Z\in\mathfrak a_{\rm AdS}\).  This is precisely the
condition required for the centrally extended bracket
\begin{equation}
 [X,Y]_{\rm ext}
 =
 [X,Y]
 +\gamma(X,Y)\,\mathbf 1
\end{equation}
to obey the Jacobi identity.

Not every two-cocycle defines a genuinely new central extension.  Given
a one-cochain
\begin{equation}
 \lambda:\mathfrak a_{\rm AdS}\longrightarrow\mathbb C ,
\end{equation}
its Chevalley-Eilenberg coboundary is, in our convention,
\begin{equation}
 (\delta\lambda)(X,Y)
 =
 -\lambda([X,Y]).
 \label{eq:two-coboundary}
\end{equation}
Two cocycles related by
\begin{equation}
 \gamma'
 =
 \gamma+\delta\lambda
\end{equation}
give equivalent central extensions, since the difference can be
absorbed into a redefinition of the generators.  Thus an element of
\(H^2\) is not an individual cocycle but an equivalence class
\begin{equation}
 [\gamma]
 =
 \gamma+B^2(\mathfrak a_{\rm AdS};\mathbb C).
\end{equation}

A simple illustration is provided by the \(LL\) sector.  The cocycle
condition for
\begin{equation}
 \gamma(L_m,L_n)
 =
 A(m)\,\delta_{m+n,0}
\end{equation}
gives
\begin{equation}
 A(m)=a_3m^3+a_1m .
 \label{eq:LL-general-solution}
\end{equation}
The two terms in \eqref{eq:LL-general-solution}, however, do not
represent two independent cohomology classes.  To see this explicitly,
take a one-cochain supported on \(L_0\),
\begin{equation}
 \lambda(L_n)=\alpha\,\delta_{n,0}.
\end{equation}
Using
\begin{equation}
 [L_m,L_n]=(m-n)L_{m+n},
\end{equation}
its coboundary is
\begin{align}
 (\delta\lambda)(L_m,L_n)
 &=
 -\lambda([L_m,L_n])
 \nonumber\\
 &=
 -(m-n)\alpha\,\delta_{m+n,0}
 \nonumber\\
 &=
 -2\alpha\,m\,\delta_{m+n,0}.
 \label{eq:LL-linear-coboundary}
\end{align}
Hence a term proportional to
\(m\,\delta_{m+n,0}\) is a two-coboundary and can be removed by changing
the representative of the cohomology class.  The linear term in
\eqref{eq:LL-general-solution} is therefore cohomologically trivial,
while the cubic term contains the nontrivial information.

We choose the standard representative
\begin{equation}
 \gamma_{LL}(L_m,L_n)
 =
 \frac{1}{12}
 m(m^2-1)\,
 \delta_{m+n,0},
 \label{eq:LL-standard-representative}
\end{equation}
so that, schematically,
\begin{equation}
 a_3m^3+a_1m
 \;\sim\;
 a_3(m^3-m),
\end{equation}
where \(\sim\) denotes equality modulo a two-coboundary.  Thus the
\(LL\) sector contributes only one independent element to
\(H^2(\mathfrak a_{\rm AdS};\mathbb C)\), namely the usual
Virasoro-type central-extension class.

\section{Operator products and radial AdS BRST anomaly channels}
\label{app:ope-details}

This appendix collects the operator-product calculations underlying the
anomaly coefficients quoted in \cref{subsec:ads-matter-ghost-currents} and
relates the possible central poles to the corresponding terms in the square
of the BRST charge.  The purpose is twofold.  First, we make explicit why
the canonical first-order AdS realization activates only the $LL$ cocycle.
Second, we show how the four central classes allowed by the abstract
residual algebra would enter $Q_B^2$ in a general quantum realization.

\subsection*{Double contractions and the anomaly coefficients.}

The complete matter and ghost current algebras were derived in
\cref{subsec:ads-matter-ghost-currents}.  Here we isolate only the
field-independent higher-order poles, since these determine which of the
central extensions allowed by the abstract algebra are actually realized
in the canonical first-order representation.

For the matter currents
\begin{equation}
 T^X=-:P_A\partial X^A:,
 \qquad
 M^X=\frac12:P_AP^A:,
 \qquad
 S^X=P_z,
\end{equation}
the only possible nontrivial double contraction occurs in
$T^X(w)T^X(v)$.  Contracting the momentum in either stress tensor with the
derivative of $X$ in the other gives
\begin{equation}
 T^X(w)T^X(v)\big|_{\rm double}
 =
 \frac{D}{(w-v)^4}.
 \label{eq:app-TT-matter-double}
\end{equation}
With the convention
\begin{equation}
 T(w)T(v)
 \sim
 \frac{c_{LL}/2}{(w-v)^4}+\cdots,
\end{equation}
this gives
\begin{equation}
 c_{LL}^X=2D.
\end{equation}

The absence of the remaining matter central terms follows directly from
the structure of the currents.  Since $M^X$ contains only momenta, the
product $T^X(w)M^X(v)$ admits at most a single contraction and therefore
cannot generate a fourth-order $LM$ pole.  Likewise, $S^X=P_z$ is linear
in the elementary fields, so $T^X(w)S^X(v)$ cannot produce the
field-independent third-order pole required for an $LS$ cocycle.
Finally, $P_z(w)P_z(v)$ is regular, excluding an $SS$ level.  Hence
\begin{equation}
 (c_{LL},c_{LM},c_{LS},c_{SS})_X
 =
 (2D,0,0,0).
 \label{eq:app-matter-anomaly-vector}
\end{equation}

The same reasoning applies to the ghost realization.  The diagonal ghost
stress tensor consists of two weight-$(2,-1)$ fermionic first-order
systems and one weight-$(1,0)$ system, whose total Virasoro contribution is
\begin{equation}
 c_{LL}^{\rm gh}
 =
 -26-26-2
 =
 -54.
 \label{eq:app-ghost-cLL}
\end{equation}
The currents associated with the remaining two constraints can be chosen as
\begin{equation}
 M^{\rm gh}
 =
 :c\partial\widetilde b:
 +2:(\partial c)\widetilde b:,
 \qquad
 S^{\rm gh}
 =
 \partial:c\beta:.
 \label{eq:app-ghost-MS-currents}
\end{equation}

Their field content excludes the additional central poles.  A central term
in $T^{\rm gh}(w)M^{\rm gh}(v)$ would require a double contraction with
both fields appearing in $M^{\rm gh}$.  These are $c$ and
$\widetilde b$, whose conjugates belong to different terms of the diagonal
ghost stress tensor; no single bilinear in $T^{\rm gh}$ can contract with
both simultaneously.  Thus no fourth-order $LM$ pole is generated.

Similarly, $S^{\rm gh}$ contains the fields $c$ and $\beta$, belonging to
the independent $(b,c)$ and $(\beta,\gamma)$ systems.  No term in the
diagonal stress tensor contains both conjugate fields $b$ and $\gamma$,
so a double contraction producing an $LS$ third-order pole is impossible.
Moreover, $S^{\rm gh}$ has no self-contraction, since neither $c$ with $c$
nor $\beta$ with $\beta$ is singular.  Therefore
\begin{equation}
 c_{LM}^{\rm gh}
 =
 c_{LS}^{\rm gh}
 =
 c_{SS}^{\rm gh}
 =
 0,
\end{equation}
and
\begin{equation}
 (c_{LL},c_{LM},c_{LS},c_{SS})_{\rm gh}
 =
 (-54,0,0,0).
 \label{eq:app-ghost-anomaly-vector}
\end{equation}

Adding the independent matter and ghost contributions gives
\begin{equation}
 (c_{LL},c_{LM},c_{LS},c_{SS})_{\rm tot}
 =
 (2D-54,0,0,0).
 \label{eq:app-total-anomaly-vector}
\end{equation}

The significance of this result is representational rather than
cohomological.  The residual algebra admits four independent central
classes, but the canonical first-order matter--ghost system considered
here realizes only the $LL$ class.  The vanishing of the $LM$, $LS$, and
$SS$ coefficients therefore reflects the structure of this particular
free-field realization and does not imply the absence of the corresponding
classes from the abstract Lie-algebra cohomology.

\subsection*{From OPE poles to mode-space cocycles.}

The numerical normalizations used for the cocycles in~\cref{app:adscohomology} follow naturally from the Laurent-mode
conventions
\begin{equation}
 T(w)=\sum_{m\in\mathbb Z}L_mw^{-m-2},
 \qquad
 M(w)=\sum_{m\in\mathbb Z}M_mw^{-m-2},
 \qquad
 S(w)=\sum_{m\in\mathbb Z}S_mw^{-m-1}.
 \label{eq:app-current-mode-expansions}
\end{equation}

Consider first a fourth-order pole
\begin{equation}
 T(w)T(v)
 \sim
 \frac{c_{LL}/2}{(w-v)^4}.
\end{equation}
The inner contour gives
\begin{equation}
 {\rm Res}_{w=v}
 \frac{w^{m+1}}{(w-v)^4}
 =
 \frac1{3!}\,
 \partial_v^3v^{m+1}
 =
 \frac{m(m^2-1)}6\,v^{m-2}.
\end{equation}
After the second contour integral one obtains
\begin{equation}
 [L_m,L_n]_{\rm central}
 =
 \frac{c_{LL}}{12}
 m(m^2-1)\delta_{m+n,0}.
 \label{eq:app-LL-mode-central}
\end{equation}
The same calculation applies to a fourth-order $T(w)M(v)$ pole normalized
as $c_{LM}/[2(w-v)^4]$, giving
\begin{equation}
 [L_m,M_n]_{\rm central}
 =
 \frac{c_{LM}}{12}
 m(m^2-1)\delta_{m+n,0}.
 \label{eq:app-LM-mode-central}
\end{equation}

For the mixed Virasoro-current channel,
\begin{equation}
 T(w)S(v)
 \sim
 \frac{c_{LS}}{(w-v)^3},
\end{equation}
and
\begin{equation}
 {\rm Res}_{w=v}
 \frac{w^{m+1}}{(w-v)^3}
 =
 \frac1{2!}\partial_v^2v^{m+1}
 =
 \frac12m(m+1)v^{m-1}.
\end{equation}
Thus
\begin{equation}
 [L_m,S_n]_{\rm central}
 =
 \frac{c_{LS}}2
 m(m+1)\delta_{m+n,0}.
 \label{eq:app-LS-mode-central}
\end{equation}

Finally,
\begin{equation}
 S(w)S(v)
 \sim
 \frac{c_{SS}}{(w-v)^2}
\end{equation}
gives
\begin{equation}
 {\rm Res}_{w=v}
 \frac{w^m}{(w-v)^2}
 =
 m v^{m-1},
\end{equation}
and hence
\begin{equation}
 [S_m,S_n]_{\rm central}
 =
 c_{SS}\,m\,\delta_{m+n,0}.
 \label{eq:app-SS-mode-central}
\end{equation}

Equations~\eqref{eq:app-LL-mode-central}--\eqref{eq:app-SS-mode-central}
explain the conventional factors $1/12$, $1/12$, $1/2$, and $1$ used for
the four representatives in~\cref{app:adscohomology}.  The
Chevalley-Eilenberg analysis fixes the allowed cocycle classes; these
numerical factors fix their normalization relative to the OPE
coefficients.

\subsection*{Central extensions in the BRST square.}

The relation between the central extensions and BRST nilpotency follows
directly from the BFV construction.  Let the quantum constraint algebra be
written schematically as
\begin{equation}
 [G_\alpha,G_\beta]
 =
 f_{\alpha\beta}{}^\delta G_\delta
 +
 K_{\alpha\beta}\,\mathbf 1 .
 \label{eq:app-quantum-constraint-algebra}
\end{equation}
The contribution of the central term to the square of the BRST charge is
\begin{equation}
 Q_B^2\big|_{\rm central}
 =
 \frac12
 C^\alpha C^\beta K_{\alpha\beta}.
 \label{eq:app-general-central-Qsquare}
\end{equation}
The noncentral part cancels against the cubic ghost terms by the ordinary
BFV mechanism and the Jacobi identities; the central part has no
constraint generator on which the remaining ghost terms can act and
therefore survives.

Using the mode conventions above, the four possible anomaly channels give
\begin{align}
 Q_B^2\big|_{\rm central}
 ={}&
 \frac{c_{LL}}{24}
 \sum_{m\in\mathbb Z}
 m(m^2-1)c_{-m}c_m
 \nonumber\\
 &+
 \frac{c_{LM}}{12}
 \sum_{m\in\mathbb Z}
 m(m^2-1)c_{-m}\widetilde c_m
 \nonumber\\
 &+
 \frac{c_{LS}}{2}
 \sum_{m\in\mathbb Z}
 m(m+1)c_{-m}\gamma_m
 \nonumber\\
 &+
 \frac{c_{SS}}{2}
 \sum_{m\in\mathbb Z}
 m\,\gamma_{-m}\gamma_m
 +
 \Delta_{\rm lin}.
 \label{eq:full-anomaly-bilinears}
\end{align}
Here $\Delta_{\rm lin}$ denotes terms associated with the linear,
representative-dependent parts of the central polynomials, equivalently
with possible normal-ordering shifts of the zero-mode constraints.  Such
terms belong to the coboundary sector and must be fixed separately from the
nontrivial cocycles displayed explicitly above.

The four ghost bilinears in
\eqref{eq:full-anomaly-bilinears} have different species content and are
linearly independent operators.  Consequently, in a general representation
BRST nilpotency requires the coefficient of every realized nontrivial
central class to vanish separately:
\begin{equation}
 c_{LL}=c_{LM}=c_{LS}=c_{SS}=0,
 \label{eq:app-general-nilpotency-central}
\end{equation}
together with the separate conditions contained in $\Delta_{\rm lin}$.

For the canonical first-order AdS realization of the main text,
\begin{equation}
 c_{LM}=c_{LS}=c_{SS}=0,
\end{equation}
identically, so that
\begin{equation}
 Q_B^2\big|_{\rm central}
 =
 \frac{2D-54}{24}
 \sum_{m\in\mathbb Z}
 m(m^2-1)c_{-m}c_m
 +
 \Delta_{\rm lin}.
\end{equation}
The nontrivial cocycle obstruction therefore reduces to the single
Virasoro-type condition
\begin{equation}
 2D-54=0,
\end{equation}
which gives $D=27$.  This conclusion concerns the nontrivial central
cohomology of the specified highest-weight representation; the remaining
linear or zero-mode conditions must still be imposed separately for
$Q_B^2=0$ as a complete operator identity.

\bibliographystyle{JHEP}
\bibliography{CW_AdS}

@article{Schild1977,
  author = {Schild, Alfred},
  title = {Classical Null Strings},
  journal = {Phys. Rev. D},
  volume = {16},
  pages = {1722},
  year = {1977},
  doi = {10.1103/PhysRevD.16.1722}
}

@article{IsbergLindstromSundborg1992,
  author = {Isberg, J. and Lindstr{\"o}m, U. and Sundborg, B.},
  title = {Space-Time Symmetries of Quantized Tensionless Strings},
  journal = {Phys. Lett. B},
  volume = {293},
  pages = {321--326},
  year = {1992},
  eprint = {hep-th/9207005},
  archivePrefix = {arXiv}
}

@article{IsbergEtAl1994,
  author = {Isberg, J. and Lindstr{\"o}m, U. and Sundborg, B. and Theodoridis, G.},
  title = {Classical and Quantized Tensionless Strings},
  journal = {Nucl. Phys. B},
  volume = {411},
  pages = {122--156},
  year = {1994},
  eprint = {hep-th/9307108},
  archivePrefix = {arXiv}
}

@article{BagchiEtAlReview,
  author = {Bagchi, Arjun and Banerjee, Aritra and Chatterjee, Ritankar and Pandit, Priyadarshini},
  title = {The Tensionless Lives of Null Strings},
  journal = {Phys. Rept.},
  volume = {1185},
  pages = {1--91},
  year = {2026},
  eprint = {2601.20959},
  archivePrefix = {arXiv},
  primaryClass = {hep-th}
}

@article{BagchiEtAlOpenNull,
  author = {Bagchi, Arjun and Chakraborty, Pronoy and Chakrabortty, Shankhadeep and Fredenhagen, Stefan and Grumiller, Daniel and Pandit, Priyadarshini},
  title = {Boundary {Carrollian} Conformal Field Theories and Open Null Strings},
  journal = {Phys. Rev. Lett.},
  volume = {134},
  pages = {071604},
  year = {2025},
  doi = {10.1103/PhysRevLett.134.071604},
  eprint = {2409.01094},
  archivePrefix = {arXiv},
  primaryClass = {hep-th}
}

@article{SheikhJabbariYavartanooOverlooked,
  author = {Sheikh-Jabbari, M. M. and Yavartanoo, H.},
  title = {An Inconsistency in the Null Strings Literature: The Tale of an Overlooked Symmetry},
  year = {2026},
  eprint = {2605.12414},
  archivePrefix = {arXiv},
  primaryClass = {hep-th}
}

@article{SheikhJabbariYavartanooI,
  author = {Sheikh-Jabbari, M. M. and Yavartanoo, H.},
  title = {Null Strings Gauged and Reloaded, {I}: Null Strings Have {Carroll-Weyl} Gauge Symmetry},
  journal = {Phys. Lett. B},
  volume = {880},
  pages = {140824},
  year = {2026},
  doi = {10.1016/j.physletb.2026.140824},
  eprint = {2605.25817},
  archivePrefix = {arXiv},
  primaryClass = {hep-th}
}

@article{SheikhJabbariYavartanooII,
  author = {Sheikh-Jabbari, M. M. and Yavartanoo, H.},
  title = {Null Strings Gauged and Reloaded, {II}: Consistent Classical Treatment of the Null Strings},
  journal = {Phys. Lett. B},
  volume = {880},
  pages = {140775},
  year = {2026},
  doi = {10.1016/j.physletb.2026.140775},
  eprint = {2605.26822},
  archivePrefix = {arXiv},
  primaryClass = {hep-th}
}

@article{SheikhJabbariYavartanooClassical,
  author = {Sheikh-Jabbari, M. M. and Yavartanoo, H.},
  title = {What is Classical Null String Theory?},
  year = {2026},
  eprint = {2609.04844},
  archivePrefix = {arXiv},
  primaryClass = {hep-th}
}

@article{LindstromSundborg,
  author = {Lindstr{\"o}m, Ulf and Sundborg, Bo},
  title = {{(Anti-)de Sitter} Null Strings and {Carroll-Weyl} Symmetry},
  year = {2026},
  eprint = {2609.07765},
  archivePrefix = {arXiv},
  primaryClass = {hep-th}
}

@article{DuaryMajiPI,
  author = {Duary, Sarthak and Maji, Sourav},
  title = {Path Integral Quantization of Null Bosonic Strings with {Carroll-Weyl} Ghosts},
  year = {2026},
  eprint = {2606.04999},
  archivePrefix = {arXiv},
  primaryClass = {hep-th}
}

@article{RasulianEtAlLightcone,
  author = {Rasulian, Ida M. and Sheikh-Jabbari, M. M. and Yavartanoo, H.},
  title = {Null-strings Gauged, Reloaded and Quantized, I: Canonical Quantization in the Light-Cone Gauge},
  year = {2026},
  eprint = {2607.02970},
  archivePrefix = {arXiv},
  primaryClass = {hep-th}
}

@article{DuaryMajiBRST,
  author = {Duary, Sarthak and Maji, Sourav},
  title = {{BRST} Quantization of {Carroll-Weyl} Gauged Null Strings},
  year = {2026},
  eprint = {2608.02731},
  archivePrefix = {arXiv},
  primaryClass = {hep-th}
}

@article{ChenHuAnomalies,
  author = {Chen, Bin and Hu, Zezhou},
  title = {Quantum Anomalies of Tensionless Bosonic Strings},
  year = {2026},
  eprint = {2608.02987},
  archivePrefix = {arXiv},
  primaryClass = {hep-th}
}

@article{NullHolography,
  author = {Sheikh-Jabbari, M. M. and Yavartanoo, H.},
  title = {Null String Holography, Null Strings Probe Projective Boundary of Their Target Spaces},
  year = {2026},
  eprint = {2609.10432},
  archivePrefix = {arXiv},
  primaryClass = {hep-th}
}

@article{BagchiEtAlScattering,
  author = {Bagchi, Arjun and Grover, Sachin and Iyer, Sharang Rajesh and Saha, Amartya and Stieberger, Stephan},
  title = {Scattering of Null Strings -- Flipped Vacuum {\&} {CHY}},
  year = {2026},
  eprint = {2609.07845},
  archivePrefix = {arXiv},
  primaryClass = {hep-th}
}

@article{BagchiWorldsheet2015,
  author = {Bagchi, Arjun and Chakrabortty, Shankhadeep and Parekh, Pulastya},
  title = {Tensionless Strings from Worldsheet Symmetries},
  journal = {JHEP},
  volume = {01},
  pages = {158},
  year = {2016},
  doi = {10.1007/JHEP01(2016)158},
  eprint = {1507.04361},
  archivePrefix = {arXiv},
  primaryClass = {hep-th}
}

@article{BagchiThreeVacua2020,
  author = {Bagchi, Arjun and Banerjee, Aritra and Chakrabortty, Shankhadeep and Dutta, Sudipta and Parekh, Pulastya},
  title = {A Tale of Three. Tensionless Strings and Vacuum Structure},
  journal = {JHEP},
  volume = {04},
  pages = {061},
  year = {2020},
  doi = {10.1007/JHEP04(2020)061},
  eprint = {2001.00354},
  archivePrefix = {arXiv},
  primaryClass = {hep-th}
}

@article{MasonSkinner2014,
  author = {Mason, Lionel and Skinner, David},
  title = {Ambitwistor Strings and the Scattering Equations},
  year = {2013},
  eprint = {1311.2564},
  archivePrefix = {arXiv},
  primaryClass = {hep-th}
}

@article{CasaliTourkine2016,
  author = {Casali, Eduardo and Tourkine, Piotr},
  title = {On the Null Origin of the Ambitwistor String},
  year = {2016},
  eprint = {1606.05636},
  archivePrefix = {arXiv},
  primaryClass = {hep-th}
}

@article{Hao:2021BMS,
    author        = {Hao, Peng-Xiang and Song, Wei and Xie, Xianjin and Zhong, Yuan},
    title         = {{BMS-invariant free scalar model}},
    eprint        = {2111.04701},
    archivePrefix = {arXiv},
    primaryClass  = {hep-th},
    doi           = {10.1103/PhysRevD.105.125005},
    journal       = {Phys. Rev. D},
    volume        = {105},
    number        = {12},
    pages         = {125005},
    year          = {2022}
}

@article{Yu:2022BMSIsing,
    author        = {Yu, Zhe-Fei and Chen, Bin},
    title         = {{Free field realization of the BMS Ising model}},
    eprint        = {2211.06926},
    archivePrefix = {arXiv},
    primaryClass  = {hep-th},
    doi           = {10.1007/JHEP08(2023)116},
    journal       = {JHEP},
    volume        = {08},
    pages         = {116},
    year          = {2023}
}

@article{Liu:2026LoopBMSKM,
    author  = {Liu, Fu},
    title   = {{Loop general BMS-Kac-Moody Lie conformal algebra}},
    doi     = {10.1016/j.geomphys.2026.105775},
    journal = {J. Geom. Phys.},
    volume  = {222},
    pages   = {105775},
    year    = {2026}
}

@article{Duary:2026izp,
    author = "Duary, Sarthak and Maji, Sourav",
    title = "{Null string holography in AdS from local BRST descent}",
    eprint = "2610.10540",
    archivePrefix = "arXiv",
    primaryClass = "hep-th",
    month = "10",
    year = "2026"
}

@article{Chen:2023esw,
    author = "Chen, Bin and Hu, Zezhou and Yu, Zhe-fei and Zheng, Yu-fan",
    title = "{Path-integral quantization of tensionless (super) string}",
    eprint = "2302.05975",
    archivePrefix = "arXiv",
    primaryClass = "hep-th",
    doi = "10.1007/JHEP08(2023)133",
    journal = "JHEP",
    volume = "08",
    pages = "133",
    year = "2023"
}

@article{Ovsienko:1998TensorDensity,
    author = "Ovsienko, Valentin Yu. and Roger, Claude",
    title = "{Generalizations of Virasoro group and Virasoro algebra
              through extensions by modules of tensor-densities on $S^1$}",
    journal = "Indag. Math.",
    volume = "9",
    number = "2",
    pages = "277--288",
    year = "1998",
    doi = "10.1016/S0019-3577(98)80024-4"
}

@article{OvsienkoRoger1996,
    author = {Ovsienko, Valentin Yu. and Roger, Claude},
    title = {Extensions of the Virasoro Group and the Virasoro Algebra by Modules of Tensor Densities on $S^1$},
    journal = {Functional Analysis and Its Applications},
    volume = {30},
    number = {4},
    pages = {290--291},
    year = {1996},
    doi = {10.1007/BF02509628},
    eprint = {hep-th/9409067},
    archivePrefix = {arXiv}
}

@article{Delfante:2026ads,
    author = "Delfante, Arnaud and Markou, Chrysoula",
    title = "{Carroll-Weyl symmetries in flat holography}",
    eprint = "2609.36026",
    archivePrefix = "arXiv",
    primaryClass = "hep-th",
    month = "9",
    year = "2026"
}

\end{document}